\documentclass{article}
\usepackage{amssymb}

\usepackage{graphicx}
\usepackage{amsmath}

\newtheorem{theorem}{Theorem}
\newtheorem{acknowledgement}[theorem]{Acknowledgment}

\newtheorem{conjecture}[theorem]{Conjecture}
\newtheorem{corollary}[theorem]{Corollary}

\newtheorem{definition}[theorem]{Definition}

\newtheorem{lemma}[theorem]{Lemma}

\newtheorem{proposition}[theorem]{Proposition}
\newtheorem{remark}[theorem]{Remark}

\newenvironment{proof}[1][Proof]{\textbf{#1.} }{\ \rule{0.5em}{0.5em}}

\newcommand{\Sym}{\mbox{{\rm Sym}}}
\newcommand{\Tor}{\mbox{{\rm Tor}}}

\newcommand{\Imm}{\mbox{{\rm Im}}}

\newcommand{\pr}[1]{\frac{\partial}{\partial #1}}
\begin{document}

\title{On deformations of Yang-Mills algebras}
\author{M. Movshev\\MPI \\ Bonn  } 
\date{\today}
\maketitle

\begin{abstract}
This is a next paper from a sequel devoted to algebraic aspects of Yang-Mills theory. We undertake a study of deformation theory of Yang-Mills algebra $YM$-a ``universal solution'' of Yang-Mills equation. 
We compute  (cyclic) (co)homology of $YM$.
\end{abstract}

\section{ Introduction.}
Yang-Mills algebra $YM$  was introduced in \cite{CD}. We in \cite{MSch} rediscovered it supersymmetric version analyzing Howe-Berkovits construction \cite{Howe} \cite{Berkovits} of Yang-Mills theory using pure spinors.

The algebra $YM$ is by definition a quotient of a free Lie algebra $Free(V)$, defined over complex numbers. It is generated by a linear space $V$, equipped with a symmetric nondegenerate inner product. Fix an orthonormal basis $v_s,s=1,\dots,n$ of $V$, then the relations of $YM$ reads as
\begin{equation}\label{E:jsjhsgtw}
\sum_{s=1}^n[v_s,[v_s,v_k]]=0, k=1,\dots,n
\end{equation}

If we substitute matrices $A_s\in Mat_N$ for $v_s$ we recover Yang-Mills (YM) equations for a theory reduced to a point.

In fact one can replace $Mat_N$ by any other Lie algebra $\mathfrak{g}$ and consider homomorphisms $\rho:YM\rightarrow \mathfrak{g}$, forming ``shell surface'' $Sol=Sol(\mathfrak{g})$. One of standard choices are homomorphisms to the matrix-valued differential operators on $\mathbb{R}^n$ of order one:
\begin{equation}\label{E:jhdfhasd}
\rho(v_s)=\pr{x_s}+A_s(x)=\nabla_s
\end{equation}
 Relations \ref{E:jsjhsgtw} in this setting  reproduce a classical YM-equations on a flat $\mathbb{R}^n$. Another choice-  $\mathfrak{g}$ is equal to Weyl algebra $W$ on $n=2k$ generators. This way we recover YM theory on noncommutative affine space. Similarly we can define YM theory on a noncommutative torus. 

It is natural to consider a variable   Lie algebra $\mathfrak{g}$ as  a ``background'' and of $YM$ as ``core'' Yang-Mills.

One of the tasks of a researcher could be to understand (in)dependence of  the theory on the background, i.e. existence of a map $\gamma_{\mathfrak{g}_1,\mathfrak{g}_2}$ which identifies $\gamma_{\mathfrak{g}_1,\mathfrak{g}_2}:Sol(\mathfrak{g}_1) \rightarrow  Sol(\mathfrak{g}_2)$. In this framework we could ask about symmetries $Sol(\mathfrak{g})$,i.e the case when  $\mathfrak{g}_1=\mathfrak{g}_2$ . Suppose $\mathfrak{g}$  has a automorphism $\alpha$. Composing $\rho\in Sol$ with $\alpha$ we get a new element $\rho_1=\alpha \circ \rho$. This construction proves symmetries useful for constructing new solutions.

One may try to apply this approach to the opposite end: take an automorphism $\beta$ and define a composition $\rho_2=\rho \circ \beta\in Sol$. The advantage of this method is that it  is universal and does not depend on $\mathfrak{g}$. In this paper among other things we will investigate such symmetries.

The infinitesimal version of an automorphism is a derivation.  We will briefly outline a homological framework, which allows to analyze differentiations of  $\mathfrak{g}$ in a systematic way.
\begin{definition}\label{D:qhashdnf}
Suppose  $\mathfrak{g}$ is an arbitrary Lie algebra and $N$ is a $\mathfrak{g}$-module. There is a complex 
\begin{equation}\label{E:xhafq}
C^k( \mathfrak{g}, N)=Hom(\Lambda^k(\mathfrak{g}),N)
\end{equation}
, called Cartan-Chevalley complex. The differential $d:C^k( \mathfrak{g},N)\rightarrow C^{k+1}( \mathfrak{g},N)$ is defined by the formula:
\begin{equation}
\begin{split}
&(dc)(l_,\dots,l_{k+1})=\sum_{i=1}^{k+1}(-1)^il_ic(l_1,\dots,\hat l_i,\dots,l_{k+1})+\\
&+\sum_{i<j}(-1)^{i+j-1}c([l_i,l_j],\dots,\hat l_i,\dots,\hat l_j,\dots,l_{k+1})
\end{split}
\end{equation}
The cohomology of this complex will be denoted by $H^k(\mathfrak{g},N)$. 
\end{definition}

We will be mostly be interested in adjoint representation $N=\mathfrak{g}$ or adjoint representation in universal enveloping $U(\mathfrak{g})$.

For $k=1$ in case of $C^k( \mathfrak{g}, \mathfrak{g})$ it is easy to see that the condition $dc=0$ is $c([l_1,l_2])=[l_1,c(l_2)]+[c(l_1),l_2]$-the condition that $c\in Der(\mathfrak{g})$ is a derivation of $\mathfrak{g}$. There is a class of trivial derivations $In(\mathfrak{g})$-so called inner derivations $c_a(l)=[a,l]$. It is natural to work with a quotient $Der(\mathfrak{g})/In(\mathfrak{g})=Out( \mathfrak{g})$. The later group by definition coincides with $H^1(\mathfrak{g},\mathfrak{g})$. As an exercise the reader can check that $H^0(\mathfrak{g},\mathfrak{g})$ coincides with the center of $\mathfrak{g}$.

Any derivation of  $\mathfrak{g}$ defines a derivation of universal enveloping  $U(\mathfrak{g})$.The converse is not true. If one would like to understand derivations of $U(\mathfrak{g})$, one has to replace $N$ by $U(\mathfrak{g})$ in \ref{E:xhafq}. The groups  $H^k(\mathfrak{g},\mathfrak{g})$  are a direct summands in  $H^k(\mathfrak{g},U(\mathfrak{g}))$.

 The group $H^2(\mathfrak{g},\mathfrak{g})$ can be interpreted as a group of nonequivalent infinitesimal deformations of  $\mathfrak{g}$. Similarly to the previous paragraph the group  $H^2(\mathfrak{g},U(\mathfrak{g}))$ classifies infinitesimal deformations of $U(\mathfrak{g})$.

Our plan is to apply the above constructions to the algebra $YM$. The Lie algebra $Der(YM)$ could be interpreted as Lie algebra of background independent vector fields, tangent  to the mass shell. The Lie algebra $In(YM)$ is a Lie algebra of  background independent gauge transformations.

We proved the following proposition
\begin{proposition}\label{P:dhfggn}
If $dim V>2$ then
$H^1(YM,U(YM))=\mathbb{C}+V+\Lambda^2(V)\supset \mathbb{C}+\Lambda^2(V)=H^1(YM,YM)$. The summands can be identified with dilation, translations and Lorentz rotations.

$H^0(YM,U(YM))=\mathbb{C}$-the center of $U(YM)$ is trivial, $H^0(YM,YM)=0$

\end{proposition}

The algebra $YM$ is by no means ``generic''. Its relations  are produced from a cyclic word ${\cal L}=\sum_{i<j}[v_i,v_j][v_i,v_j]$ ( compare with $tr(\sum_{i<j}[A_i,A_j][A_i,A_j]), A_s\in Mat_N$), taking partial derivatives $\frac{\partial {\cal L}}{\partial v_s}$. One can deform ${\cal L}$ by adding some small $h{\cal L}'$. Such modification would deform the relations and hence the algebra
. Not all $h{\cal L}'$ produce a nontrivial deformation, because there is a ``field redefinition''. Similarly two different $h{\cal L}'$ and $h{\cal L}''$ could give isomorphic algebra. In such case ${\cal L}'$ and ${\cal L}''$ are called equivalent. The linear  space $L=\bigoplus L_i$ of all nonequivalent ${\cal L}'$ is graded by eigenvalues of dilation. An element $v_s$ has grading $2$. One can form a generating function $F(t)=\sum dim L_kt^k$.
\begin{proposition}
\begin{equation}\label{E:bcdshd}
\begin{split}
&F(t)=1-\sum_{k\geq 1}ln(1-dimVt^{2k}+dimVt^{4k}-t^{8k})\frac{\psi(k)}{k}+\\
&+\left( \frac{dimV(dimV-1)}{2}-1\right)t^8+dimVt^6
\end{split}
\end{equation}
Recall that
$\psi(k)=\sum_{l|k} 1$
\end{proposition}
We postpone a discussion of appearance of the last two summands in \ref{E:bcdshd} until a mathematical part of the introduction.

It is interesting to understand what is the most generic deformation of the algebra $YM$. Deformations associated with deformation of the lagrangian were discussed above. It turns out that the algebra $YM$ admits a non lagrangian deformation.

Fix a complex-valued skew-symmetric matrix $\omega=\omega^{kl}$, $1\leq k,l \leq n$ and a row of numbers $a=(a_1,\dots a_n)$.

Define an algebra $YM(\omega,a)$ as a quotient of a free algebra $T(V)$ (not a free Lie algebra!!!) by relations
\begin{equation}\label{E:ddddd}
f^j=\sum_{s=1}^n\left( [v_s,[v_s,v_j]]+a_s[v_s,v_j] \right)+\sum_{kl}\omega^{kl}\{v_k,[v_l,v_j]\},\quad j=1\dots n
\end{equation}
As usual we set the degree of $v_j$ equal to two, $\{a,b\}\overset{def}{=}ab+ba$

One can associate a differential graded algebra $bv(\omega,a)$ to  $YM(\omega,a)$: the algebra $bv(\omega,a)$ is generated by $v_j,v^{*j},c, j=1\dots n$, $degv_j=2, degv^{*j}=7, deg c=10$. The differential generalizes standard BV differential:
\begin{equation}\label{E:differ}
\begin{split}
&d(v_j)=0\\
&d(v^{*j})=f^j\\
&d(c)=\sum_{s=1}^n\left( [v_s,v^{*s}]-a_sv^{*s}\right)+\sum_{kl}\omega^{kl}\{v_k,v^{*l}\}
\end{split}
\end{equation}
It is easy to see that $d^2=0$

\begin{remark}
Due to the linear term $\sum_sa_sv^{*s}$  if formula for $d(c)$ in \ref{E:differ} the algebras with $(a_s)\neq 0$ have homological dimension equal to two.
\end{remark}

We defer details about $YM(a,\omega)$, an explicit constructions of generic deformation of $YM$ by incorporating a choice of lagrangian in $YM(\omega,a)$ construction  together with discussion of formality of $U(YM)$ to future publications.


There is one particular background we would like to discuss in more details- flat background of $\mathbb{R}^n$. We however impose no restrictions on the gauge group. It was proved in \cite{MSch2} that any solution, i.e. homomorphism of the form \ref{E:jhdfhasd} is completely determined by the traceless part of 
\begin{equation}\label{E:fahdid}
\nabla_{(i_1}\dots\nabla_{i_k}F_{i_{k+1})i_{k+2}}|_{x=0}
\end{equation}
 In the last formula $F_{st}=[\nabla_s,\nabla_t]$ is the curvature, $()$-stands for symmetrization. For this purposes we in \cite{MSch2} introduced a free Lie algebra $TYM=Free(M)$.  The linear space $M$ is graded. The $(k+2)$-nd component $M_{k+2}$ is a linear space of traceless $k+2$ tensors $m_{i_1,\dots,i_{k},i_{k+1},i_{k+2}}$, which are symmetric with respect to first $k+1$ indecies and antisymmetric with respect to the last two. Another description of $M$ is that it is a dense linear space in a Fourier  transform of gauge classes of solutions of Maxwell equation. The formula \ref{E:fahdid} gives a representation of $TYM$ in the the gauge group. It turns out that any representation of $TYM$ (see \cite{MSch2}) can be used to construct a formal solution of YM equation in a flat space.

To study symmetries which do not depend on the gauge group one has to study space of outer derivations of $Free(M)$. It is unconstrained but enormously big. The linear space $M$ is an irreducible representation of conformal Lie algebra  $\mathfrak{so}_{n+2}$ $(dimV>2)$, which contains Lorentz rotations $\mathfrak{so}_{n}$. It can be easily promoted to automorphism of $TYM=Free(M)$, giving an example of  symmetries. In dimension four conformal group is a manifest group of symmetries of the lagrangian, in any other dimension it is not.

It is a nontrivial task to find a formula for a symmetry in terms of connections, rather then derivatives of connections \ref{E:fahdid}. A complication comes from non locality of the most of such symmetries. 

Polyakov in \cite{Pol} among other things considered a problem of finding a generating function $c(t)$ of dimensions of spaces of cyclic words in $U(TYM)$. He found a formula by direct counting. In our approach the formula 
\begin{equation}
\begin{split}
&c(t)=1-\sum_{k>0}ln(1-M(t^k))\frac{\psi(k)}{k}\\
&1-M(t)=\frac{1-dimVt+dimVt^3-t^4}{(1-t)^{dimV}}
\end{split}
\end{equation}
follows from general algebraic considerations (see Appendix).

From mathematical point of view algebra $YM$ is a graded  algebra with Poincare duality in (co)homology of homological dimension $3$. There is a  theory of such algebras.

 Denote $HH^{\bullet}(A,A)$-Hochschild cohomology, $HH_{\bullet}(A,A)$-Hochschild homology. $HC_{\bullet}(A)$-cyclic homology (in our case $A$ becomes $U(YM)$, $HH_{\bullet}(A,A)$ can be substituted by  $H_{\bullet}(YM,U(YM))$ and  $HH^{\bullet}(A,A)$ by $ H^{\bullet}(YM,U(YM))$). Details about cyclic homology can be found in \cite{Loday}.  It is a general fact that for such algebras $HH^{i}(A,A)=HH_{3-i}(A,A)$  and 
\begin{equation}\label{E:sghxiehc}
0 \rightarrow \overline{HC}_{i-1}(A) \rightarrow \overline{HH}_{i}(A,A) \rightarrow \overline{HC}_{i}(A) \rightarrow 0
\end{equation}
The over-line symbol  denotes reduced theories. There is also an isomorphism $\overline{HH}_{i}(A,A)=HH_{i}(A,A), i>0$. 
See \cite{CD} for application of these ideas to $U(YM)$.

There is one example of algebras of such type where deformation theory is understood. These are so called Sklyanin algebras. Being deformations of polynomial algebras on three variables, these algebras have a polynomial growth. There is a deformation spectral sequence which allows to estimate Hochschild cohomology groups of deformed algebra, by Hochschild cohomology of  the undeformed. This idea was employed by Van der Berg in \cite{VDB} .

The $YM$ algebras  have exponential growth (see section \ref{SS:sjshd}). They are  more of a kin to  free algebras and by no means are deformations of polynomials.  In this paper we developed new methods to deal with Hochschild cohomology of such algebras.

Let us briefly outline the main steps in computation of (cyclic) (co)homology of $U(YM)$. Some of the preliminary consideration can also be found in \cite{CD}, which were discovered  the author independently. 

Due to the short exact sequence \ref{E:sghxiehc} and vanishing of $HH_i(YM,U(YM)), i\geq 4$, we conclude that $\overline{HC}_i(U(YM))=0,i\geq 3$. Thus we have $\overline{HC}_2(U(YM))=HH_3(YM,U(YM))$. We have isomorphisms 
\begin{equation}
HH_3(YM,U(YM))=HH^3(YM,U(YM))=Z(U(YM))\mbox{ - the center}
\end{equation}
Thus $\overline{HC}_2(U(YM))=Z(U(YM))$. We have an isomorphism $$HH_2(YM,U(YM))=HH^1(YM,U(YM))=Out(U(YM))$$. The later group contains a derivation $eu$ corresponding to the grading. It defines an inclusion $a\rightarrow  eu\cup a$ of $Z(U(YM)$ into $Out(U(YM)$, which splits  projection  
\begin{equation}
\overline{HC}_1(U(YM)) \rightarrow  HH_2(YM,U(YM))\overset{p}{\rightarrow} \overline{HC}_2(U(YM))=Z(U(YM))
\end{equation}
Thus we can identify 
\begin{equation}
\begin{split}
&\overline{HC}_2(U(YM))=Z(U(YM))\\
&\overline{HC}_1(U(YM))=Out(U(YM))/Z(U(YM))\\
&\overline{HC}_0(U(YM))=\overline{HH}_0(YM,U(YM))
\end{split}
\end{equation}
From this we con completely recover the group content of $HH_{\bullet}(YM,U(YM))$ and $HH^{\bullet}(YM,U(YM))$:
\begin{equation}\label{E:dfsnbvcsx}
\begin{split}
&HH^0(YM,U(YM))=Z(U(YM))\mbox{ -tautology }\\
&HH^1(YM,U(YM))=Out(U(YM))\mbox{ -tautology }\\
&HH^2(YM,U(YM))=Out(U(YM))/Z(U(YM))+HH_0(YM,U(YM))/\mathbb{C}\\
&HH^3(YM,U(YM))=HH_0(YM,U(YM))\\
\end{split}
\end{equation}
The map $HH^1(YM,U(YM))\rightarrow HH^2(YM,U(YM))$  used in \ref{E:dfsnbvcsx} is defined by the formula $b\rightarrow eu\cup b$. 

It is worthwhile to point out that the deformations constructed in \ref{E:ddddd}are governed by cocycles $Out(U(YM))/Z(U(YM))\subset HH^2(YM,U(YM))$ 

We can conclude that in order to know the group content of \\ $HH^{\bullet}(YM,U(YM))$ it suffice to know $\overline{HC}_{\bullet}(U(YM))$.

For a graded algebra reduced cyclic homology is graded $\overline{HC}_{i}(U(YM))=\bigoplus_{j\geq 0}\overline{HC}_{ij}(U(YM))$. One can form a generating function $\chi(t)$ of Euler characteristics:
\begin{equation}
\chi(t)=\sum_{i=0}^2\sum_{j\geq0}(-1)^idim(\overline{HC}_{ij}(U(YM)))t^j
\end{equation}
There is an explicit formula for $\chi(t)$ in terms of generating function $U(YM)(t)=\sum_{i\geq 0}dimU(YM)_it^i=\frac{1}{1-dimVt^2+dimVt^6-t^8}$,  proved in Appendix.  The formula for $\chi(t)$:
\begin{equation}
\chi(t)=-\sum_{k\geq 1}ln(1-dimVt^{2k}+dimVt^{4k}-t^{8k})\frac{\psi(k)}{k}
\end{equation}
is also proved in Appendix.
 If we knew  dimensions $\overline{HC}_{ij}$ for $i=1,2$, we would have had a formula for generating function $\overline{HC}_{i}(i)=\sum_{j\geq0}dim(\overline{HC}_{ij}(U(YM)))t^j$, $i=0$ and hence for all other values.

This explains our interest to  $Z(U(YM))$ and $Out(U(YM))$. It worthwhile to notice that nontrivially of these groups is responsibly for extra two summands in \ref{E:bcdshd}.  It seems that general homological  consideration are not sufficient to find the groups in question  and new ideas are needed.

The main idea, which stays behind all our computation is that Lie algebra $YM=\bigoplus_{i\geq 2}YM_i$ contains a free subalgebra ( an ideal ) $TYM=\bigoplus_{i> 2}YM_i$ (see \cite{MSch}).

We compute  the groups $H^i(YM,U(TYM)), i=0,1$ first. To do this we introduce a filtration $F^i$ of $U(TYM)$ generated by powers of augmentation ideal in $U(TYM)$. It gives a spectral sequence, converging to $H^{\bullet}(YM,U(TYM))$, whose $E^{ij}_2$ term is isomorphic to $H^i(YM,M^{\otimes j})$, where the module $M$ has been described above . 

The most difficult task was to compute cohomology of the higher differential of the mentioned spectral sequence (see section \ref{S:wjwudu}, \ref{S:iewgafvf}), which enables us to prove that $H^0(YM,U(TYM))=\mathbb{C}$ and $H^1(YM,U(TYM))=V+V$ (see section \ref{S:wuiehjdb}).

Having made this crucial step we can step over computations $H^{i}(YM,U(YM))$ (see section \ref{S:oquebnx}). We replace $U(YM)$ by $\Sym(YM)$, because these representations are isomorphic. We introduce a filtration on $\Sym(YM)$ defined by powers of ideal, generated by linear space $TYM$. This filtration provides us with a spectral sequence. We estimate in $E_2$ term that the groups $H^{i}(YM,U(YM))$ are not larger then stated in proposition \ref{P:dhfggn}. The estimate from below is obvious.

The paper is organized as follows:
In section \ref{S:ensnsjs} we setup some notations.

In section \ref{S:hwtdjkg} we accumulate some vanishing results needed in section \ref{S:wuiehjdb}.

In section \ref{S:wyxcnfiabnfo} is mathematically most interesting part. We show an existence of recursive relations between cohomology $H^{\bullet}(YM,M^{\otimes j})$ (the $YM$-module $M$ was defined in the introduction). The relations has a similarity with Connes exact sequence.

In section \ref{S:rusdjcn} we prove that $H^{1}(YM,M^{\otimes j})=0$ $j>1$.

In section \ref{S:eiwcdnds} we undertake a detailed study of module $M$ and groups $\Tor(M,M)$. The results will be used in section \ref{S:wuiehjdb}.



In section \ref{S:wuiehjdb} we use our knowledge obtained in sections \ref{S:hwtdjkg} and \ref{S:eiwcdnds} to compute $H^i(YM,U(TYM)), \quad i=0,1$. At this point we are on midway to the proof of proposition \ref{P:dhfggn}.

Section \ref{S:wjwudu}, \ref{S:iewgafvf} are technically most difficult. We compute some segments of a spectral sequence introduced in \ref{S:wuiehjdb}. 

In section \ref{S:oquebnx} is a culmination. We prove proposition \ref{P:dhfggn}.


\begin{acknowledgement}
The  author would like to thank IHES and MPI, where the most of the work has been done. He also would like to thank  M.Kontsevich,N.Nekrasov, A.S.Schwarz,  D.Sullivan for useful discussions, M.Rocek for opportunity to present this material at "2005 Simons workshop".
\end{acknowledgement}
\section{Notations}\label{S:ensnsjs}

$\mathbb{C}$ stands for a field of complex numbers or a trivial representation of a Lie algebra.

$Free(W)$ stands for a free Lie algebra on linear space $W$

$T(W)$ - free associative algebra.

$\Sym(W)$ -  free commutative algebra ( algebra of polynomials).

$\Lambda(W)$ -  free anticommutative (Grassman) algebra.

A linear space $V$(complexified flat space-time) will be used throughout this paper, $dim(V)=n$. This space is equipped with nondegenerate symmetric bilinear form. To simplify the formulas we will typically choose an orthonormal basis in the space $V$. It enables us to make no distinction between upper and lower tensor indecies.

Let $\mathfrak{g}$ be a Lie algebra. By $U(\mathfrak{g})$ we denote the universal enveloping of $\mathfrak{g}$.

There is an analog of cyclic homology for Lie algebras. Recall that for an algebra $A$ the group $HC_0(A)$ is the group of cotraces $A/[A,A]$. For a  Lie algebra  $\mathfrak{g}$ the analog of a trace $tr(a)$ is $\mathfrak{g}$-invariant dot product. 
\begin{definition}\label{D:whysbxg}
For a Lie algebra $\mathfrak{g}$ the group $D(\mathfrak{g})$ is a group of inner $\mathfrak{g}$-invariant co-products:
 $D(\mathfrak{g})=\Sym^2(\mathfrak{g})_{\mathfrak{g}}$. The linear space $D(\mathfrak{g})$ is generated by elements $a\circ b, a,b\in\mathfrak{g}$. Subject to relation $[a,b]\circ c+b\circ [a,c]=0, a\circ b=b\circ a$ and the symbol is linear with respect to each of the arguments.
\end{definition}
\section{Vanishing results}\label{S:hwtdjkg}
As it was mentioned in the introduction the proof of the main proposition \ref{P:dhfggn} uses spectral sequence technique. One is ingredients of the proof is to show that certain terms of the spectral sequence vanish. This is proved in current section. The main result in this section is lemma \ref{P:hsbxb}.

\subsection{Generalities about cohomology of $YM$}

Let us set notations: $v_i, i=1\dots n$ stand for generators of $YM$, $x_i, i=1\dots n$ stand for generators of it abelenization $V$. 

Suppose $W$ is $YM$ module. The Lie algebra cohomology groups $H^{\bullet}(YM,W)$ were defined in the introduction (see \ref{D:qhashdnf}), the  homology $H_{\bullet}(YM,W)$ are similar and  defined in \cite{CE}.
According to \cite{MSch2} there is an alternative way to compute $H^i(YM,W)=H_{3-i}(YM,W)$.
Construct a complex 
\begin{equation}\label{nnasf}
C(W)=0\rightarrow W\overset {d_0}{\rightarrow}W\otimes V\overset {d_1}{\rightarrow}W\otimes V\overset {d_2}{\rightarrow}W\rightarrow 0
\end{equation}
 with differentials given by the formulas:

\begin{align}
&d_0w=\sum_{1\leq s \leq n}\rho(v_s)w\otimes A^{*s}\label{E:uwnbs0}\\
&d_1w\otimes A^{*i}=\sum _{1\leq s \leq n}( \rho(v^2_s)w\otimes A_i- 2\rho(v_iv_s)w\otimes A_s+\rho(v_sv_i)w\otimes A_s)\label{E:uwnbs1}\\
&d_2w\otimes A_i=\rho(v_i)w\label{E:uwnbs2}
\end{align}
where $\rho:U(YM)\rightarrow End(W)$ is representation.
\begin{proposition}\cite{MSch2}
$H^i(YM,W)=H^i(C(W))$
\end{proposition}

The Lie algebra $YM$ contains a Lie ideal, which consists of elements of degree $\geq 3$. We denote this ideal by $TYM$. As usual we denote $U(\mathfrak{g})$ the universal enveloping algebra of Lie algebra $\mathfrak{g}$.

One of the examples of representation s of $YM$ is $\Sym(V)$-the universal enveloping of abelenization of $YM$. It admits two commuting actions of $YM$: left and right multiplications, (they in fact coincide). 
\begin{proposition}\label{P:cxbvcvg}\cite{MSch2}

 The second homology of $C(\Sym(V))$ is nontrivial and equal to $M$- right $YM$ module (it can be transformed to left module by the standard trick); the action of $YM$ on $M$ factors through $V$.

$H^3C(\Sym(V))=\mathbb{C}$, all other cohomology vanish.
\end{proposition}

It was proved in \cite{MSch2} that $TYM$ is a free Lie algebra.
The space of its generators coincides with a linear space $M$, which is $\Sym(V)$-module. 


%
The universal enveloping algebra of a free Lie algebra $Free(X)$ is a free associative algebra $T(X)$. These two objects, though belong to two different categories, have the same space of generators $X$. An application of this observation  to our setup is that $U(TYM)$ is isomorphic to $T(M)$. If we fix an infinite basis $<e_1,\dots,e_s,\dots>$ of $M$ then any  element of $U(TYM)$ can be uniquely represented as a linear combination of monomials $e_{i_1}\dots e_{i_k}$

The action of $YM$ on $U(TYM)$ is nonlinear in a sense that $x_im=m_1+\sum m_2^{'i}m^{''i}_2+\dots $, where $m, m_1,m_2^{'i},m^{''i}_2,\dots $ are elements of generating space  $M$, $x_i$ is a generator of degree two in $YM$

One can simplify the structure of the action by discarding nonlinear terms starting with $m^{'i}_2 m^{''i}_2$. Mathematically speaking we have $U(TYM)=\bigoplus_{i\geq 0}M^{\otimes j}$ and the action of $YM$ preserves a filtration $F^pU(TYM)=\bigoplus_{j\geq p}M^{\otimes j}$(in the above constructions tensor product is taken over complex numbers). A certain simplification of action can be achieved through adjoint quotient construction $GrU(TYM)$. The action of $YM$ preserves the additional grading  $Gr^jU(TYM)$. In particular $YM$ acts on the generating space $Gr^1U(TYM)=M$.
%

%

One can recover  the action of $YM$ on $GrU(TYM)=\bigoplus_{j\geq 0}M^{\otimes j}$ from action on $M$ using Leibniz rule.

One can use the complex \ref{nnasf} to compute cohomology $H^i(YM,U(TYM))$ and of $H^i(YM,GrU(TYM))$. The later groups can be considered as an approximation of the former. As  $H^i(YM,GrU(TYM))=\bigoplus_{j\geq 0}H^i(YM,M^{\otimes j})$ and since it will be possible to prove vanishing of some of  $H^i(YM,M^{\otimes j})$ we will concentrate on $Gr$-version of coefficients of cohomology.

\subsection{On long exact sequence for  $H^i(YM,M^{\otimes j})$}\label{S:wyxcnfiabnfo}
The most important technical tool used in this paper - long exact sequence \ref{E:focnmd} is established in this section. We will use it throughout the paper.

Denote $\Lambda(V,N)=\Lambda(V)\otimes N$ -the homological  Cartan-Chevalley complex of abelian Lie algebra $V$ with coefficient in module $N$ . Since $V$ is abelian, it is equal to Koszul complex of $\Sym(V)$- module $N$. Denote the cohomology of $\Lambda(V,N)$ by $H(V,N)$. Denote $\varsigma_i$ generators of $\Lambda(V)$

One can notice some simplification in in the form of differential $d_1$ (\ref{E:uwnbs1}) for the module $M^{\otimes j}$.
Since the action of $YM$ factors through abelian algebra, we have $$d_1w\otimes A^{*i}=\sum _{1\leq s\leq n}(x_s^2w\otimes A_i- x_sx_iw\otimes A_s)$$ which coincides with the formula (\ref{E:uwnbs1}) for $d_1$ in the complex (\ref{nnasf}).

 We write differential in $C(M^{\otimes j})$ explicitly for later references. Denote $w_1|\dots|w_j$ a monomial in $M^{\otimes j}$.

\begin{align}
& d_0w_1|\dots|w_j=\sum_{1\leq s \leq n}\sum_{1\leq k \leq j}w_1|\dots |x_iw_k|\dots|w_j\otimes A^{*i}\\
&d_1w_1|\dots|w_j\otimes A^{*i}=\sum _{1\leq s\leq n}\sum _{1\leq k,l\leq j }(w_1|\dots |x_s w_l|\dots|x_s w_k|\dots |w_n\otimes A_i -\\
&- w_1| x_s w_k\dots| x_iw_l|\dots| w_j\otimes A_s)\\
&d_2w_1|\dots| w_j \otimes x_i=\sum_{1\leq k \leq j}w|\dots| x_iw_k|\dots|w_j
\end{align}

%
For any commutative algebra $C$ and a module $B$ there is a canonical identification of modules $C\underset{C}{\otimes} B=B$. We would like to specialize this construction to the case of $C=\Sym(V)$, $B=M^{\otimes j}$. Such specialization has its own specifics: the algebra $\Sym(V)$ is a universal enveloping algebra of abelian Lie algebra $V$. An isomorphism $M^{\otimes j}\underset{\Sym(V)}{\otimes}\Sym(V)$ can be formulated in terms of coinvariants.

\begin{definition}
For any Lie algebra $\mathfrak{g}$ and a $\mathfrak{g}$-module $H$ denote $H_{\mathfrak{g}}$ a quotient space $H/\mathfrak{g}H$.
\end{definition}

There are two commuting structures $V$ module on $M^{\otimes j}\otimes \Sym(V)$.
The first one is $v(a\otimes b)=va\otimes b-a\otimes vb$ where $a\otimes b\in M^{\otimes j}\otimes \Sym(V)$.
The second one is $va\otimes b=a\otimes vb$.

It obvious that $(M^{\otimes j}\otimes \Sym(V))_V$ with respect to the first structure is nothing else but $M^{\otimes j}\underset{\Sym(V)}{\otimes} \Sym(V)$. It coincides with $\Sym(V)$ module $M^{\otimes j}$ with respect to the remaining second module structure.
%
\begin{proposition}\label{P:odfhsfwr}
There is an exact sequence
\begin{equation}\label{E:focnmd}
\begin{split}
&0\rightarrow H_3(V,M^{\otimes j})\overset{S_j}{\rightarrow} H_1(V,M^{\otimes (j+1)})\overset{B_j}{\rightarrow} H_2(YM,M^{\otimes j})\overset{I_j}{\rightarrow}\\
& \rightarrow H_2(V,M^{\otimes j})\overset{S_j}{\rightarrow} H_0(V,M^{\otimes (j+1)})\overset{B_j}{\rightarrow} H_1(YM,M^{\otimes j})\overset{I_j}{\rightarrow} H_1(V,M^{\otimes j})\rightarrow 0 
\end{split}
\end{equation}

and isomorphisms
\begin{equation}
H_3(YM,M^{\otimes j})=0 \quad j \geq 1
\end{equation}
\begin{equation}\label{E:bnxgsetygf}
H_0(YM,M^{\otimes j})=H_0(V,M^{\otimes j})
\end{equation}
\begin{equation}\label{E:bsye}
H_s(V,M^{\otimes j})=\Lambda^{2j+s}[V]\quad s\geq 2, \mbox{ except } s=2,j=1
\end{equation}
For $j=1$ we have
\begin{equation}\label{E:shgscsh}
\begin{split}
&0\rightarrow \Lambda ^4V\rightarrow H_2(V,M)\rightarrow \mathbb {C}\rightarrow 0\\
&0\rightarrow \Lambda ^3V\rightarrow H_1(V,M)\rightarrow V\rightarrow 0\\
&\quad \Lambda ^2V\widetilde {\rightarrow}H_0(V,M)
\end{split}
\end{equation}
There is $\Sym(V)$-linear map 
\begin{equation}
\delta^c:H_{i}(YM,M^{\otimes j})\rightarrow H_{i+1}(YM,M^{\otimes j-1})
\end{equation}

Denote composition $B_{j-1}\circ I_j=\delta^c_j$. Then $\delta^c_{j-1}\circ \delta^c_j=0$

\end{proposition}
\begin{proof}
Due to general homological arguments there is a quasiisomorphism 
\begin{equation}\label{POkkl3}
C(M^{\otimes j})\cong C(\Lambda (V,M^{\otimes j}\otimes \Sym (V))=\Lambda (V,M^{\otimes j}\otimes C(\Sym (V)),
\end{equation}
where $\cong $ is a quasiisomorphism and $=$ is an isomorphism.

Indeed we can replace a tensor product $M^{\otimes j}=M^{\otimes j}\underset{\Sym(V)}{\otimes}\Sym(V)$ by homotopical version of it- $\Lambda(V,M^{\otimes j}\otimes\Sym(V))$. This is a valid procedure (in a sense that cohomology of $\Lambda(V,M^{\otimes j}\otimes\Sym(V))$ coincide with $M^{\otimes j}$  because $\Sym(V)$ is  a free cyclic module over $\Sym(V)$).

There is an isomorphism of complexes $\Lambda(V,C(M^{\otimes j}\otimes\Sym(V))=\Lambda(V,M^{\otimes j}\otimes C(\Sym(V))$, because the differential in $C(M^{\otimes j}\otimes\Sym(V)$ does not depend on $M^{\otimes j}$-factor.

The complex $\Lambda(V,M^{\otimes j}\otimes C(\Sym(V))$ has  a filtration 
\begin{equation}
F^i\Lambda(V,M^{\otimes j}\otimes C(\Sym(V))=\bigoplus_{s\leq i}\Lambda^s[V]\otimes M^{\otimes j}\otimes C(\Sym(V))
\end{equation}

The $E^2$ term of the corresponding spectral sequence is equal to 
\begin{equation}
\begin{split}
&E^2_{2,i}=H_i(V,M^{\otimes (j+1)}) \mbox{ because } H^2(C(\Sym(V))=M  \mbox{ (prop. \ref{P:cxbvcvg})}\\
&E^2_{3,i}=H_i(V,M^{\otimes j})\mbox{ because } H^3(C(\Sym(V))=\mathbb{C} \mbox{ (prop. \ref{P:cxbvcvg})}
\end{split}
\end{equation}
The only possible higher differential $\delta:E^2_{2,i}\rightarrow E^2_{3,i-2}$ defines a map $H_{i+2}(V,M^{\otimes j})\overset{\delta}\rightarrow H_{i}(V,M^{\otimes (j+1)})$. The spectral sequence converges to $H^{\bullet}(C(M^{\otimes j}))\overset{def}{=}H_{\bullet}(YM,M^{\otimes j})$. Thus we can rewrite the spectral sequence in a form of long exact sequence:

\begin{equation}\label{E:cvshyf}
\begin{split}
&\dots \rightarrow H_i(YM,M^{\otimes j}) \rightarrow H_i(V,M^{\otimes j})\overset{\delta}{\rightarrow} H_{i-2}(V,M^{\otimes (j+1)})\rightarrow \\
&\rightarrow H_{i-1}(YM,M^{\otimes j})\rightarrow \dots
\end{split}
\end{equation}
We can make use of some obvious vanishing restrictions: $H_i(V,M^{\otimes j})=0$ for $i<0$ and for $i> dimV$. It was proved in \cite{MSch2} that the algebra $YM$ has homological dimension $3$ and as a result $H_i(YM,M^{\otimes j})=0$ for $i>3$ and $i<0$.  Moreover for any $j>0$ an operator of multiplication on $\sum_s a_sx_s$ has no kernel( formally this is because a product of $j$ quadrics is a smooth irreducible algebraic manifold and $M^{\otimes j}$ is a space of sections of appropriate vector bundle over it). Thus $H_3(YM,M^{\otimes j})=0$. 

This information is enough to deduce exact sequence \ref{E:focnmd} and isomorphism \ref{E:bnxgsetygf} from \ref{E:cvshyf}
restricting later to small values of $i$. For large values of $i$ due to vanishing of $H^i(YM,M^{\otimes j})$, we have isomorphisms:
\begin{equation}
H_i(V,M^{\otimes (j+1)})=H_{i+2}(V,M^{\otimes j})\mbox{ for } j>0\quad i>2 \mbox{ and  for }j=0\quad i>2
\end{equation}
Since $H_{i}(V,\mathbb{C})=\Lambda^iV$ we conclude that \ref{E:bsye} holds. 
This enables us to write exact sequence \ref{E:focnmd} in its final form.
We can apply \ref{E:cvshyf} to $j=0$ case. This way we obtain \ref{E:shgscsh} 
%
\end{proof}

\begin{proposition}\label{P:iuxjxs}
The map $S_1$ is an embedding.
\end{proposition}
\begin{proof}
The cocycles in $H_2(V,M)=\Lambda^2(V)+\mathbb{C}$ are spanned by $F_{[ij}\otimes \varsigma_k \wedge \varsigma_{l]}$ and $\sum_{i<j}F_{ij}\otimes \varsigma_i \wedge \varsigma_{j}$. The $S_1$ images of these cocycles in $H_0(V,M^{\otimes 2})$ are $F_{[ij}\otimes F_{kl]}$ and $\sum_{i<j}F_{ij}\otimes F_{ij}$ which are clearly linearly independent (by grading reasons) elements of $H_0(V,M^{\otimes 2})$.
\end{proof}
\begin{corollary}\label{C:vcksfgfw}
The map $B_2$ is surjective
\end{corollary}
\begin{proof}
Combine  propositions  \ref{P:iuxjxs} and \ref{P:odfhsfwr}.
\end{proof}
\subsection{Proof of the vanishing lemma}\label{S:rusdjcn}
Consider a module $N$ over algebra $\Sym(V)$, $V=<x_1,\dots,x_n>$, $q=\sum_{i=1}^n x^2$.
\begin{definition}\label{D:bwgxn}
Denote $Ann(N)=\{m\in N|am=0 \mbox{ for }a\in \Sym(V)\mbox{ such that } a(0)=0\}$.

Define $Z(N)=\{m_i\in N, i=1\dots n|\sum_{i=1}^nx_kx_im_i=qm_k\}$, $B(N)=\{m_i=x_im|m\in N \}$. It is easy to see that $B(N)\subset Z(N)$. Denote $H(N)=Z(N)/B(N)$.
Define a map $p:H(N)\rightarrow Ann(N/qN)$ by the formula 
\begin{equation}\label{E:nbfskgf}
\nu:m_1,\dots,m_n\rightarrow \sum_ix_im_i
\end{equation}
\end{definition}
\begin{proposition}\label{P:fsjvf}
If multiplication on $q$ in $N$ has no kernel then the map $\nu$ is an isomorphism.
\end{proposition}

\begin{proof}
We check that the map $\nu$ (\ref{E:nbfskgf}) is  correctly defined first.
Since $x_k\sum_{i=1}^nx_im_i=qm_k\in qN$ the element $\sum_{i=1}^nx_im_i\in Ann(N/qN)$. The element $m_i=x_in$ maps into $qn\in qN$ is identically zero in $Ann(N/qN)$.

Suppose $m\in Ann(N/qN)$. Then by definition of $Ann(N/qN)$ we can find some elements $m_i\in N$ such that $x_im=qm_i$ . It implies that $\sum_ix_iqm_i=qm$ and $q\nu(m_1,\dots,m_n)=qm$.  Assume that multiplication on $q$ has a trivial kernel. Then we can cancel on $q$. It proves that map $\nu$ is surjective.

Suppose $\nu(m_1,\dots,m_n)=0\in Ann(N/qN)$. It means that there is an element $m\in N$ such that $qm=\sum_ix_im_i$.By assumption  $x_k\sum_ix_im_i=qm_k$. Hence we have $x_kqm=qm_k$. After canceling on $q$ we see that $(m_1,\dots,m_n)$ is a trivial element in $H(N)$.
\end{proof}

It is easy to interpret the group $Ann(N)$ in terms of Koszul complex $\Lambda(V,N)$. A direct inspection shows that a map $m\rightarrow mA_1\wedge\dots\wedge A_n$ defines an isomorphism of groups $Ann(N)\cong H_n(V,N)$.

\begin{proposition}\label{P:nhcgs}
Suppose a multiplication on element $q$ in module $N$ has no kernel and  $H_n(V,N)=H_{n-1}(V,N)=0$. Then the group $H(N)$ is trivial.
\end{proposition}
\begin{proof}
Due to assumptions we have a short exact sequence $0\rightarrow N\overset{q}{\rightarrow}N\rightarrow N/qN\rightarrow 0$. It produces   a long exact sequence, whose terminal segment will of interest to us: $$0\rightarrow H_n(V,N)\rightarrow H_n(V,N)\rightarrow H_n(V,N/qN)\rightarrow H_{n-1}(V,N)\rightarrow \dots$$
Due to vanishing of $H_{n-1}(V,N)$ and $H_{n-1}(V,N)$  we have $H_n(V,N/qN)=0$. The proposition \ref{P:fsjvf} implies that than $H(N)=0$.
\end{proof}

We would like to apply proposition \ref{P:nhcgs} to modules $M^{\otimes j}$. To do that we need to check that multiplication on $q$ has no kernel. 

It is convenient to introduce the following notations.
\begin{definition}
The algebra $\Sym(V)$, generated by $x_1,\dots,x_n$ has a homomorphism $\Delta:\Sym(V)\rightarrow \Sym(V)\otimes\Sym(V)$, defined by the formula $\Delta(x_i)=x_i\otimes 1+1\otimes x_i$. This homomorphism is call a diagonal. It can be used to define a map $\Delta^2\Sym(V)\rightarrow \Sym(V)^{\otimes 3}$, by the formula $(\Delta \otimes 1)\circ\Delta=(1\otimes\Delta) \circ\Delta$. There are similar maps $\Delta^{j-1}$ to $\Sym(V)^{\otimes j}$, which also do not depend of a way we present them as composition of $\Delta$.
\end{definition}

It is easy to see that element $q$ acts on $M^{\otimes j}$ through multiplication on the image of $\Delta^{j-1}(q)$ in $A^{\otimes j}$. Though the image of $q$ in $A$ is trivial, the image of $q$ in $A^{\otimes 2}$ is equal to $2\sum_sx_s\otimes x_s$. More generally we have 
\begin{equation}\label{E:jsgxs}
\begin{split}
&\Delta^{j-1}(q)=\Delta^{j-3}(q)\otimes 1^{\otimes 2}+\sum_{1\leq l\leq j-2}\sum_s1\otimes \dots \overset{l}{x}_s\dots\otimes 1\otimes(x_s\otimes 1+1\otimes x_s)+\\
&+1^{\otimes j}\otimes\sum_sx_s\otimes x_s=a+b+c
\end{split}
\end{equation}
The element $\overset{l}{x}_s$ stands in the $l$-th position in the tensor product.
\begin{proposition}\label{P:jdhcb}
Operator of multiplication on $\Delta^{j-1}(q)$ in $M^{\otimes j}$ has no kernel for $j\geq 2$.
\end{proposition}
\begin{proof}

The module $M=\bigoplus_{i\geq 0}M_i$ is graded, the module $M^{\otimes j}$ is polygraded. A subscript in a homogeneous element $m_{i_1,\dots,i_j}$ denotes polygrading of module component the element comes from.
We will probe this statement for $j=2$ first. Without loss of generality we may assume that element $m=\sum_{i+j=k}m_{i,j}$ is of total degree $k$. If $\sum_sx_s\otimes x_sm=0$ then vanishing must happen for every individual summand, because $q$ for $j=2$ does not mix the components. The element $m_{i,j}$ is an element of $H^0(Q^{\times 2},{\cal M}(i)\boxtimes {\cal M}(j))$, where ${\cal M}$ is some homogeneous vector bundle on quadric $Q$. Similarly $\sum_sx_s\otimes x_s\in H^0(Q^{\times 2},{\cal O}(1)\boxtimes {\cal O}(1))$. Having this interpretation of elements $\sum_sx_s\otimes x_s$ and $m_{ij}$ we see that identity $\sum_sx_s\otimes x_sm_{ij}$ must hold pointvise over $Q^{\times 2}$. For sections of locally free shaves over smooth algebraic manifold this implies that one of the section must be zero.  It  is $m_{ij}$ in our case.

Let us now turn to a general case $j>2$. In the module $M^{\otimes (j-2)}$ introduce a filtration $F^pM^{\otimes (j-2)}=\bigoplus_{i_1+\dots i_{j-2}\geq p}M_{i_1}\otimes \dots \otimes M_{i_{j-2}}$ and a filtration $F^pM^{\otimes j}=F^pM^{\otimes (j-2)}\otimes M^{\otimes 2}$.
\begin{lemma}

1. The operator of multiplication on $\Delta^{j-1}(q)$ preserves filtration $F^pM^{\otimes j}$.

2. On $GrM^{\otimes j}$ the action of multiplication on $\Delta^{j-1}(q)$ coincides with multiplication on element $c$ from \ref{E:jsgxs} and has no kernel.

\end{lemma}
\begin{proof}
The elements $a,b,c$ are defined in \ref{E:jsgxs}.
By definition we have the following inclusions: $aF^pM^{\otimes j}\subset F^{p+2}M^{\otimes j}$, $bF^pM^{\otimes j}\subset F^{p+1}M^{\otimes j}$. It implies that only $c$ acts nontrivially on $GrM^{\otimes j}$, which has no kernel by the proof for the case $j=2$.
\end{proof}

\begin{lemma}
Suppose $A,B$ are filtered finite dimensional vector spaces. A map $\psi:A\rightarrow B$ preserves filtration and defines injection $Gr\psi$ of $GrA$ into $GrB$. Then the map $\psi$ is an injection. 
\end{lemma}
\begin{proof}
This is left to the reader.
\end{proof}

In our case linear spaces $A$ and $B$ are graded components on $M^{\otimes j}$ of degree $k$ and $k+2$. The filtration is induced by filtration from $F^pM^{\otimes j}$. 

From this we conclude that operator of multiplication on $\Delta^{j-1}(q)$ has no kernel on $M^{\otimes j}$ if $j\geq 2$.
\end{proof}

\begin{lemma}\label{P:hsbxb}
For $j\geq 2$ $H^1(YM,M^{\otimes j})=0$
\end{lemma}
\begin{proof}
We can use complex $C(M^{\otimes j})$ (\ref{nnasf}) to compute such cohomology. A direct inspection shows that $H^1(C(M^{\otimes j}))$ coincides with the group $H(M^{\otimes j})$, defined in \ref{D:bwgxn}. According to proposition \ref{P:jdhcb} the operator of multiplication on $\Delta^{j-1}(q)$ in $M^{\otimes j}$ has no kernel. Then by proposition \ref{P:fsjvf} the group $H(M^{\otimes j})$ is isomorphic to $H_n(V,M^{\otimes j}/qM^{\otimes j})$. By proposition \ref{P:odfhsfwr} the groups $H_n(V,M^{\otimes j})$ and $H_{n-1}(V,M^{\otimes j})$ are equal to zero, hence by proposition \ref{P:nhcgs} we have $H(M^{\otimes j})=0$.
\end{proof}

\begin{corollary}
The module $M^{\otimes j}$ is free over $\Sym (V)$ for $j> \mbox{max}(1,\frac{n-3}{2})$ .
\end{corollary}
\begin{proof}
We use vanishing result of $H_i(V,M^{\otimes j}), i\geq 1$ from proposition \ref{P:hsbxb} in conjunction with proposition \ref{P:odfhsfwr}, provided $j> \mbox{max}(1,\frac{n-3}{2})$. This condition is necessary and sufficient for a graded module  $M^{\otimes j}$  to be free.
\end{proof}

Suppose $X$ is a smooth algebraic variety, equipped with very ample line bundle ${\cal L}$. Denote by $A=\bigoplus_{i\geq 0}A_i$ the corresponding ring of homogeneous functions. Let ${\cal N}$ be an algebraic vector bundle and $N$ the corresponding $A$-module. Let $W=A_1$. Then $N$ is a module over abelian Lie algebra $W$. 
\begin{conjecture}
There is a constant $n(X,{\cal L})$ such that for any $N_i$ ($i=1,\dots,n(X,{\cal L})$) as above the module $N_1\underset{\mathbb{C}}{\otimes} N_2 \underset{\mathbb{C}}{\otimes}\dots \underset{\mathbb{C}}{\otimes}N_{n(X,{\cal L})}$ is free over $\Sym(W)$ - universal enveloping of $W$.
\end{conjecture}

\section{Homological properties of the module $M$}\label{S:eiwcdnds}
As the reader can see lemma \ref{P:hsbxb} gives us information about $H^1(YM,M^{\otimes j})$ for $j>1$. In our computation of cohomology $H(YM,U(YM))$ we will need to have a good understanding of $H^1(YM,M)$, which is by no means is zero. As we shall see it is an infinite-dimensional space. Since the groups $H_{\bullet}(YM,M)$ and $H_{\bullet}(V,M\otimes M)$ are intimately related through exact sequence \ref{E:focnmd}, we will concentrate on the later group.	The main proposition of this section is \ref{P:vsbfh}, where  $H_{\bullet}(V,M\otimes M)$ are computed. This result will be used in the proof of proposition \ref{P:ududufgqex}.

For a pair $\Sym(V)$-modules $N,K$ the cohomology $H_{i,j}(V,N\otimes K)$ (the second index comes from the module grading) computes groups $\Tor^{\Sym(V)}_{i,j}(N,K)=\Tor_{i,j}(N,K)$. We drop algebra dependence of $\Tor$, because the later be always computed over $\Sym(V)$. A group $\Tor_{i,j}(N,K)$ can be computed by other means, e.g. by use of resolutions (see \cite{CE}). This was a motivation for a use of  different notations. If only one index is present it stands for homological index.

\subsection{Structure of minimal $\Sym(V)$-resolution of module $M$}\label{S:uehcmgo}
The following fact is standard in homological algebra: for any graded $\Sym(V)$-module $N$ a minimal free resolution has a form
\begin{equation}\label{E:xhsgc}
N\overset{d_{-1}}{\leftarrow} \Tor_0(N,\mathbb{C})\otimes \Sym(V)\overset{d_{0}}{\leftarrow}\dots \overset{d_{n-1}}{\leftarrow} \Tor_n(N,\mathbb{C})\otimes \Sym(V)\leftarrow 0=R(N)
\end{equation}

The  grading that exists on modules can be lifted to $\Tor$ groups, making it a bigraded group $\Tor_i(N,K)=\bigoplus_{j\geq i}\Tor_{i,j}(N,K)$.


\begin{proposition}
$\Tor_i(M,\mathbb{C})=\Tor_{ii}(M,\mathbb{C})$.
\end{proposition}
\begin{remark}
A graded $\Sym(V)$-module $N$ for which $deg(\Tor_i(M,\mathbb{C}))=i$ is called Koszul.
\end{remark}
We can recover the differential in \ref{E:xhsgc} from homogeneity and $SO(n)$-invariance.

The group $\Tor_i(M,\mathbb{C})$ contains a regular part $\Lambda^{2+i}(V)$ and two sporadic pieces: $\tilde{V}=V\subset \Tor_1(M,\mathbb{C})$ and $<e>=\mathbb{C}\subset \Tor_2(M,\mathbb{C})$. Denote $<e_i>, i=1,\dots,n$ a basis of $V$,$<f_i>, i=1,\dots,n$ a basis of $\tilde{V}$.
\begin{proposition}
Restriction of the differential $d$ on regular part of $R(M)$ coincides with Koszul differential 
\begin{equation}
d(ae_{i_1}\wedge\dots\wedge e_{i_k})=\sum_s ax_{i_s}(-1)^se_{i_1}\wedge\dots \hat{e}_{i_s}\dots \wedge e_{i_k}
\end{equation}
where $a\in \Sym(V)$.
Restriction of the differential on the exceptional part is:
\begin{equation}
\begin{split}
&d(af_i)=a\sum_jx_je_j\wedge e_i\\
&d(ae)=a\sum_j f_jx_j
\end{split}
\end{equation}

\end{proposition}
\begin{proof}
We leave to the reader a proof that above formulas are the only possible $SO(n)$ equivariant nontrivial $\Sym(V)$ maps between modules $R_i(M)$ and $R_{i-1}(M)$. 
\end{proof}

We can sheafify the complex $R(M)$ by replacing it by direct sum of complexes of sheaves on $\mathbf{P}^{n-1}$ of the form:
\begin{equation}\label{E:jsvxyf}
{\cal M}(i)\overset{d_{-1}}{\leftarrow} \Tor_0(M,\mathbb{C})(i)\overset{d_{0}}{\leftarrow}\dots \overset{d_{n-1}}{\leftarrow} \Tor_n(M,\mathbb{C})(i-n)\leftarrow 0=R({\cal M}(i))
\end{equation}
\begin{proposition}
The complex of sheaves \ref{E:jsvxyf} is acyclic
\end{proposition}
\begin{proof}
This is a simple corollary of Serre equivalence between category coherent sheaves on $\mathbf{P}^{n-1}$ and a category of finitely generated graded  $\Sym(V)$-modules, modulo finite-dimensional modules.
\end{proof}

\subsection{On the structure of $\Tor_{ij}(M,M)$}\label{S:ewurncf}

We will be dealing with  the complex \ref{E:jsvxyf} restricted on a nonsingular  quadric $Q=supp({\cal M}(i))$, in particular we will use its sheaf(local) cohomology.

To get a better understanding of such restriction introduce a $\Sym(V)$ module $A=\Sym(V)/q\Sym(V)$, where $q=\sum_sx^2_s=0$ is a homogeneous equation of the quadric $Q$.

Denote $i:Q\rightarrow \mathbf{P}^{n-1}$ the inclusion, $i_*{\cal O}$ the direct image of the structure sheaf of $Q$.

We would like to compute local(sheaf) cohomology of the complex 
\begin{equation}\label{E:dcvxjkff}
\Tor_0(M,\mathbb{C})(i)\underset{{\cal O}}{\otimes}i_*{\cal O}\overset{d_{0}}{\leftarrow}\dots \overset{d_{n-1}}{\leftarrow} \Tor_n(M,\mathbb{C})(i-n)\underset{{\cal O}}{\otimes}i_*{\cal O} 
\end{equation} 
The later is completely determined by cohomology of 
\begin{equation}\label{E:vhjfrgvxj}
\Tor_0(M,\mathbb{C})\otimes A_i\overset{d_{0}}{\leftarrow}\dots \overset{d_{n-1}}{\leftarrow} \Tor_n(M,\mathbb{C})\otimes A_{i-n}
\end{equation}for large values of $i$ (use Serre equivalence).

It turns out that the later groups can be computed for all values of $i$, because the cohomology of \ref{E:vhjfrgvxj} is $\Tor_i(M,A)$. These can be computed by resolving the second argument. Indeed there is a resolution
\begin{equation}\label{E:cvgsdhsk}
A\leftarrow \Sym(V)\overset{ \times q}{\leftarrow}\Sym(V)
\end{equation}
A generator of the  first free cyclic $\Sym(V)$-module lives in degree zero, while a generator of the  second in degree two(because the quadric has degree two). 

The complex $\Lambda(V,A\otimes A)$ computes groups $\Tor(A,A)$. The later due to existence of resolution \ref{E:cvgsdhsk} are all equal to zero, except $\Tor_{0,i}(A,A)=A_i, \Tor_{1,i}(A,A)=A_{i-2}$.
\begin{proposition}\label{P:gener}
In the complex $\Lambda(V,A\otimes A)$ a generator of $\Tor_{1,2}(A,A)$  over $A$ is equal to 
\begin{equation}
\sum_{s=1}^n(x_s\otimes1-1\otimes x_s )\otimes\varsigma_s
\end{equation}
\end{proposition}
\begin{proof}
Easy check.
\end{proof}

After tensor multiplication \ref{E:cvgsdhsk} on $M$ over $\Sym(V)$ we obtain a complex with trivial differential (multiplication in $M$ on $q$ is identically equal to zero).
\begin{proposition}\label{P:hsczc}
The following identity folds: $\Tor_{0,i}(M,A)=M_i$, $\Tor_{1,i}(M,A)=M_{i-2}$, $\Tor_{k,i}(M,A)=0,k\geq 2$.
\end{proposition}

It is possible to represent classes $\Tor_{1,i}(M,A)$ by explicit cocycles in $\Lambda(V,M\otimes A)$(we do not bother to represent all classes, because we will be dealing only with the later group).

\begin{proposition}\label{P:sffxsdhjc}
There is a map $\psi:M\rightarrow M\otimes A\otimes \Lambda^1[V]$, defined by the formula
\begin{equation}
\psi(m)=\sum_{s=1}^n(x_sm\otimes1-m\otimes x_s)\otimes \varsigma_s
\end{equation}
It satisfies $d\psi(m)=0$ and induced map $M\rightarrow H_1(V,M\otimes A)$ is an isomorphism.
\end{proposition}
\begin{proof}
Left to the reader. As a hint we say that an essential part of the proof is a spectral sequence associated with filtration of $M$, defined by the grading.
\end{proof}

As a corollary of  proposition \ref{P:hsczc} we obtain
\begin{proposition}
The local cohomology of the complex \ref{E:dcvxjkff} is equal to ${\cal M}(i)$ in zero degree and ${\cal M}(i-2)$ in degree one.
\end{proposition}

The knowledge we gained in this section  enables us to compute local cohomology of the complex
\begin{equation}\label{E:vedvxjkff}
\Tor_0(M,\mathbb{C})\otimes {\cal M}(i)\overset{d_{0}}{\leftarrow}\dots \overset{d_{n-1}}{\leftarrow} \Tor_n(M,\mathbb{C})\otimes{\cal M}(i-n)
\end{equation}
obtained by tensoring on \ref{E:dcvxjkff} on ${\cal M}$. 
\begin{proposition}
Local cohomology of \ref{E:vedvxjkff} is equal to ${\cal M}^{\otimes 2}(i)$ in degree zero and ${\cal M}^{\otimes 2}(i-2)$ in degree one.
\end{proposition}

\begin{proof}
The analysis is purely local over $Q$. It has no difference with computation for \ref{E:dcvxjkff} because locally ${\cal M}$ is isomorphic to ${\cal O}^{\oplus k}$ for some $k$. 
\end{proof}

\begin{definition}
Denote $[w_1,w_2,\dots,w_{[n/2]}]$ an irreducible representation of $\mathfrak{so}_n$, which has the highest weight with coordinated $[w_1,w_2,\dots,w_{[n/2]}]$ in a standardly ordered basis of fundamental weights. For example $[1,0,\dots,0]$ corresponds to defining representation in linear space $V$, $[0,\dots,1,\dots,0]$ where $1$ stands in the i-th place corresponds to $\Lambda^i(V)$ ($i\leq n/2$), $[0,2,\dots,0]$ is the largest irreducible representation in $\Sym^2(\Lambda^2(V))$ and so on.
\end{definition}

\begin{proposition}\label{P:vsbfh}
\begin{equation}
\mbox{
\scriptsize{
$\begin{array}{l|l|l}
\begin{array}{l}\Tor_{0,0}(M,M)=\\\left[0,2,\dots,0\right]+\mathbb{C}\\\left[2,0,\dots,0\right]+\Lambda^2(V)\\\left[1,0,1,\dots,0\right]+\Lambda^4(V)\end{array}&&  \\\hline
 \begin{array}{l}\Tor_{0,1}(M,M)= \\\left[1,2,\dots,0\right]\\\left[3,0,\dots,0\right]\\\left[2,0,1,\dots,0\right]\end{array}& &\begin{array}{l}\Tor_{1,1}(M,M)=\\V+\\ \Lambda^3(V)+\Lambda^5(V)\end{array} \\\hline
\begin{array}{l}\Tor_{0,2}(M,M)=\\\left[2,2,\dots,0\right]+\\\left[4,0,\dots,0\right]+\\\left[3,0,1,\dots,0\right]\end{array}&\begin{array}{l}\Tor_{1,2}(M,M)=\\\left[0,2,\dots,0\right]+\\\left[2,0,\dots,0\right]+\\\left[1,0,1,\dots,0\right]\end{array} &\begin{array}{l} \Tor_{2,2}(M,M)=\\\Lambda^6(V)\end{array} \\\hline
&\dots&  \\\hline
\begin{array}{l}\Tor_{0,i}(M,M)=\\\left[i,2,\dots,0\right]+\\\left[i+2,0,\dots,0\right]+\\\left[i+1,0,1,\dots,0\right]\end{array}& \begin{array}{l}\Tor_{1,i}(M,M)=\\\left[i-2,2,\dots,0\right]+\\\left[i,0,\dots,0\right]+\\\left[i-1,0,1,\dots,0\right]\end{array} &\begin{array}{l}\Tor_{i,i}(M,M)=\\\Lambda^{i+4}(V)\end{array} \\\hline
&\dots& 
\end{array}$
}
}
\end{equation}
\end{proposition}
\begin{proof}

There are two spectral sequences converging to hypercohomology of the complex \ref{E:vedvxjkff}. The first one has $E_1$ term isomorphic to $E_1^{s,-t}=\Tor_s(M,\mathbb{C})\otimes H^t(Q,{\cal M}(i-s))$. Observe that for $t=0$ we recover the  complex
\begin{equation}\label{E:qfagcyd}
\Tor_0(M,\mathbb{C})\otimes M_{i}\overset{d_{0}}{\leftarrow}\dots \overset{d_{n-1}}{\leftarrow} \Tor_n(M,\mathbb{C})\otimes M_{i-n}
\end{equation}
This complex by  definition from \cite{McL} is $R(M)\underset{\Sym(V)}{\otimes}M$, should compute the groups $\Tor_{ij}(M,M)$.

The cohomology of the sheaves ${\cal M}(i)$ are 
\begin{equation}
\begin{split}
&H^0(Q,{\cal M}(i))=M_i,\quad i\geq 0\\
&H^1(Q,{\cal M}(-2))=\mathbb{C}\\
&H^{n-3}(Q,{\cal M}(2-n))=\mathbb{C}\\
&H^{n-2}(Q,{\cal M}(-n-i))=M^*_i,\quad i\geq 0
\end{split}
\end{equation}

The groups were computed using Borel-Weyl-Bott theorem.

For $t>0$ almost all groups $E_1^{s,-t}$ are equal to zero, making computations particularly nice. Exceptional entries are $(s,t)=(i+2,-1)$. In this case $E_1^{s,-t}$ is equal to $\Tor_{i+2}(M,\mathbb{C})$-the groups computed in proposition \ref{E:focnmd}.

Another exception is $(s,t)=(n-2,-(n-3))$, $E_1^{s,-t}=\Tor_{n-2}(M,\mathbb{C})=\mathbb{C}$.

This gives us a good control over discrepancy between the limiting term of the spectral sequence and cohomology of \ref{E:qfagcyd}.

The second spectral sequence has second term $\tilde E^{s,-t}_2$ equal to zero, except $s=0$ and $\tilde E^{0,-t}_2=H^t(Q,{\cal M}^{\otimes 2}(i))$ and $s=1$,$\tilde E^{1,-t}_2=H^t(Q,{\cal M}^{\otimes 2}(i-2))$. This spectral sequence degenerated in $\tilde E_2$-term, due to representation-theoretic considerations.
The sheaf ${\cal M}$ is isomorphic to the tangent bundle to the quadric $Q$ (see \cite{MSch2} for explanation). The Picard group of $Q$ is isomorphic to $\mathbb{Z}$, with an ample generator ${\cal O}(1)$. The canonical class is equal to $K={\cal O}(2-n)$. There is a direct sum decomposition ${\cal M}^{\otimes 2}=\Sym^2({\cal M})+\Lambda^2({\cal M})$. Moreover there is a decomposition $\Sym^2({\cal M})={\cal O}(2)+\Sym_{irr}^2({\cal M})$. The projection $\Sym^2({\cal M})\rightarrow {\cal O}(2)$ defines a nondegenerate symmetric pairing on ${\cal M}$, which enables to identify ${\cal M}(-2)\cong {\cal M}^*$

The following tables (computed again  using Borel-Weyl-Bott theorem) give representation content of cohomologies of $\Lambda^2({\cal M})(i)$, $\Sym_{irr}^2({\cal M})(i)$ and ${\cal O}(i)$:
\begin{equation}
\begin{split}
&H^0(Q,\Sym_{irr}^2({\cal M})(i))=[i,2,\dots,0]\quad i\geq 0\\
&H^1(Q,\Sym_{irr}^2({\cal M})(-2))=[0,1,\dots,0]\\
&H^1(Q,\Sym_{irr}^2({\cal M})(-3))=[1,0,\dots,0]\\
&H^{n-3}(Q,\Sym_{irr}^2({\cal M})(-n+1))=[1,0,\dots,0]\\
&H^{n-3}(Q,\Sym_{irr}^2({\cal M})(-n+2))=[0,1,\dots,0]\\
&H^{n-2}(Q,\Sym_{irr}^2({\cal M})(-n-i-2))=[i,2,\dots,0]\\
\end{split}
\end{equation}
\begin{equation}
\begin{split}
&H^0(Q,\Lambda^2({\cal M})(i))=[i,0,1,\dots,0]\quad i\geq 0\\
&H^2(Q,\Lambda^2({\cal M})(-3))=[0,0,\dots,0]\\
&H^{n-4}(Q,\Lambda^2({\cal M})(3-n))=[0,0,\dots,0]\\
&H^{n-2}(Q,\Lambda^2({\cal M})(-n-i))=[i,0,1\dots,0]\\
\end{split}
\end{equation}
\begin{equation}
\begin{split}
&H^0(Q,{\cal O}(i))=[i,0,0,\dots,0]\quad i\geq 0\\
&H^{n-2}(Q,\Lambda^2({\cal M})(2-n-i))=[i,0,0\dots,0]\\
\end{split}
\end{equation}

There is an isomorphism $\Lambda^2(V)\otimes \Lambda^2(V)=[0,1,0,\dots,0]\otimes[0,1,0,\dots,0]=\Sym^2(V)+\Lambda^2(V)+\Lambda^4(V)+[1,0,1\dots,0]+[0,2,\dots,0]$

There is an isomorphism
\begin{equation}
\begin{split}
&H^0(Q,{\cal O}(i))=[i,0,0,\dots,0]\quad i\geq 0\\
&H^{n-2}(Q,\Lambda^2({\cal M})(2-n-i))=[i,0,0\dots,0]\\
\end{split} 
\end{equation}
Finally
\begin{equation}
\begin{split}
&H^0(Q,{\cal M}^{\otimes 2}(-2))=\mathbb{C}\quad H^1(Q,{\cal M}^{\otimes 2}(-2))=\Lambda^2(V)\\
&H^0(Q,{\cal M}^{\otimes 2}(-1))=V+\Lambda^3(V)\\
&H^0(Q,{\cal M}^{\otimes 2}(i))=[i,2,\dots,0]+[i+1,0,1\dots,0]+[i+2,0,\dots,0]\quad i\geq 0
\end{split} 
\end{equation}

The second spectral sequence $\tilde E_r$ indicates that hypercohomology of the complex \ref{E:vedvxjkff} live in degrees $0,1$, $i\geq 1$. It makes all higher differential $d_r, r\geq 3$ in the first  spectral sequence $ E_r$ nontrivial. It means that the groups $\Tor_{i+2}(M.\mathbb{C})$ contribute (via appropriate differential) to cohomology of \ref{E:qfagcyd} .

Since we are working over a field the cohomology of \ref{E:qfagcyd} is a direct sum of hypercohomology of \ref{E:vedvxjkff} and groups  $\Tor_{\bullet}(M.\mathbb{C})$.

The case $i=0$ can be computed by hands.

We hope that the interested reader will be able to fill in all missing details
\end{proof}

Suppose $C,N$ are an algebra and a $C$-module. Choose elements $x_1,\dots,x_n$. It makes $C,N$ into $\Sym(V)$-modules. We can consider complexes $\Lambda(V,C)$ and $\Lambda(V,N)$. The action $C\otimes N\rightarrow N$ induces a map of complexes $\Lambda(V,C)\otimes \Lambda(V,N) \rightarrow \Lambda(V,N)$, which induces a map of cohomology $\times H^i(V,C)\otimes H^j(V,N) \rightarrow H^{i+j}(V,N)$.

We can specialize this construction to $C=A\otimes A$, $N=M\otimes M$. The elements are $xs\otimes1+1\otimes x_s, s=1\dots n$
We have a map 
\begin{equation}\label{P:jshsgsf}
\times H_1(V,A\otimes A)\otimes H_0(V,M\otimes M)\rightarrow H_1(V,M\otimes M)
\end{equation}
\begin{proposition}\label{P:twrfsxkd}
The cokernel of the map \ref{P:jshsgsf} is equal to $\Lambda^5(V)+\Lambda^3(V)+V$
\end{proposition}
\begin{proof}
The proof mimics the proof of proposition \ref{P:vsbfh}. There are analogs of both spectral sequences for complexes $\Lambda(V,A\otimes A)$ and $ \Lambda(V,M\otimes M)$. On the spectral sequences there is a multiplicative structure corresponding to $\times$. Consider second terms of the second spectral sequences . These are equal to $\tilde E_2^{0,-t}(A)=H^t(Q,{\cal O}(i)), \tilde E_2^{1,-t}(A)=H^t(Q,{\cal O}(i-2)$; $\tilde E_2^{0,-t}(M)=H^t(Q,{\cal M}^{\otimes 2}(i), \tilde E_2^{1,-t}(M)=H^t(Q,{\cal M}^{\otimes 2}(i-2)$( the $i$-th graded component). The map 
\begin{equation}
\tilde E_2^{1,0}(A)\otimes \tilde E_2^{0,0}(M)=H^0(Q,{\cal O}(i-2))\otimes H^0(Q,{\cal M}^{\otimes 2}(j))\rightarrow H^0(Q,{\cal M}(i+j-2))=\tilde E_2^{1,0}(M)
\end{equation}
 is surjective in the following cases. Suppose $i+j-2=k$, then $H^0(Q,{\cal M}(k))$ is in range if $j\geq 0 $ and $i-2\geq 2$. The minimal $k$ of this form is equal to zero. Thus we see that the only nontrivial group not within the range is $H^0(Q,{\cal M}(k)), k=-1$. The later group is equal to $\Lambda^3(V)+V$. It is easy to enumerate cocycles in  $Tor_1(M,M)$  which span the last group.

The cocycles spanning $V$ are
\begin{equation}\label{E:shsc}
\sum_{i<j}F_{ij}\circ F_{ij}\otimes \varsigma_s+2\sum_{jk}F_{sj}\circ F_{jk}\otimes \varsigma_k
\end{equation}
The cocycles spanning $\Lambda^3(V)$ are of the form
\begin{equation}\label{E:chdcs}
\sum_{s}F_{[ks}\wedge F_{ms}\otimes \varsigma_{m]}
\end{equation}
As usual we denote $\circ,\wedge$ symmetric (resp. skewsymmetric) tensor product.
Thus the multiplication is surjective on the limiting term.

Let us turn to the maps of first spectral sequences. The limiting terms of both spectral sequence coincide. It implies that  it is surjective on $E_2$-term (recall that it is equal to $H(V,A\otimes A)$ or $H(V,M\otimes M)$ plus one standalone simple group), modulo images of higher differential. These images are easy to calculate. The  higher differential maps $Tor_3(M,\mathbb{C})=\Lambda^5(V)$ into  $H_1(V,M\otimes M)$ . 

The image of $Tor_3(M,\mathbb{C})$ in $H_1(V,M\otimes M)$ is spanned by cocycles of the form
\begin{equation}\label{E:dte}
F_{[ij}\otimes F_{ks}\otimes \varsigma_{s]}
\end{equation}
as usual $[]$-sign denote  skewsymmetrization.

Independently of the above it is also quite clear that the span of elements \ref{E:dte} does not intersect   the image of the map $\times$
\end{proof}

\begin{corollary}\label{E:lfgjdw}
Up to linear combination of \ref{E:shsc},\ref{E:chdcs},\ref{E:dte} every cocycle in $H_1(V,M\otimes M)$ can be represented as
\begin{equation}
\sum_s\left(\sum_i x_sa_i\otimes b_i-a_i\otimes x_sb_i\right)\otimes \varsigma_s
\end{equation}
\end{corollary}
\begin{proof}

Combine propositions \ref{P:gener} and \ref{P:twrfsxkd}. 
\end{proof}

\begin{proposition}
The cohomology $H^1(YM,M)$ is equal to $\bigoplus_{i\geq -1}[i,2,\dots,0]+[i+2,0,\dots,0]+[i+1,0,1,\dots,0]$.
It is a module over ring $A=\Sym(V)/(q)$. The generating cocycles are
\begin{equation}
\begin{split}
&\sum_iF_{ij}\otimes A^{*}_{i}\\
&F_{[ij}\otimes A^{*}_{k]}\\
&F_{ij}x_{[k}\otimes A^{*}_{l]}+F_{kl}x_{[i}\otimes A^{*}_{j]}\\
\end{split}
\end{equation}
\end{proposition}

\section{Computation of the cohomology $H^1(YM,U(TYM))$}\label{S:wuiehjdb}
At this point we are on a mid way collecting facts for the proof of proposition \ref{P:dhfggn}. Proposition \ref{P:ududufgqex} proved in this section is the key for \ref{P:dhfggn}. However proposition \ref{P:ududufgqex} has some interest by its own, because $H_{\bullet}(YM,U(TYM))$ can be interpreted as equivariant homology with respect to the group of translations.
\begin{proposition}\label{P:ududufgqex}
Let $IU(TYM)$ be the augmentation ideal of the universal enveloping of Lie algebra $TYM$. Then
$H^0(YM,IU(TYM))=0, H^1(YM,IU(TYM))=V$
\end{proposition}

\begin{proof}
The universal enveloping algebra $U(TYM)=T(M)$ admits a filtration by powers of the augmentation ideal $I\subset U(TYM)$. The adjoint action of $YM$ preserves $I$, hence the filtration $F^i=I^{\times i}$. We plan to compute cohomology $H^i(YM,U(TYM))$ using a spectral sequence of mentioned filtration. 

The $E_2$ term of it  is equal to $E^{ij}_2=H^{i+j}(YM,M^{\otimes j})$. In the previous section we computed $H^1(YM,M^{\otimes j})$ for $j\geq 1$. According to proposition \ref{P:hsbxb}  the groups are equal to zero for $j\geq2$. Our goal is to examen  the differential in the spectral sequence on the group $H^1(YM,M)$. The differential $\delta$ acts:
\begin{equation}\label{E:kdjvsh}
\delta:H_2(YM,M)\rightarrow H_1(YM,\Lambda^2[M])\subset H_1(YM,M^{\otimes 2})
\end{equation}

In the following part of the section we shall prove that the kernel of $\delta$ is equal to $V$.
\end{proof}

Observe that there are a surjective maps $B_1:H_1(V,M^{\otimes 2})\rightarrow H_2(YM,M)$(see corollary \ref{C:vcksfgfw}) and $I_2:H_1(YM,M^{\otimes 2})\rightarrow H_1(V,M^{\otimes 2})$ (see proposition \ref{P:odfhsfwr}) from long exact sequence \ref{E:focnmd}.

 We call elements of  $H_2(YM,M)$, which belong to $\Imm(H_1(V,\Sym^2(M))$-symmetric, to $\Imm(H_1(V,\Lambda^2(M))$-antisymmetric.

\subsection{Analysis of $\delta$, restricted on antisymmetric part of $H_2(YM,M)$}\label{S:wjwudu}

Recall that conventional way to compute $H_{\bullet}(YM,M^{\otimes j})$ is through Cartan-Chevalley complex $M^{\otimes j}\otimes \Lambda(YM)$

Explicit construction of the map $B_1$ is given in the next proposition.
\begin{proposition}
Pick a Cartan-Chevalley cocycle $\sum_i a_i\otimes b_i\otimes v_i$ representing an element $x\in H_1(V,M^{\otimes 2})$. 
There is a procedure of constructing elements $c_i,\tilde c_i\in TYM$ such that an element 
\begin{equation}\label{P:jhdgtwnd}
\tilde x=\sum_i a_i\otimes (b_i\wedge v_i)+\sum_i a_i\otimes (c_i\wedge \tilde c_i) \quad c_i,\tilde c_i\in TYM
\end{equation} 
is a cocycle in $M\otimes\Lambda^2(YM)$-component of Cartan-Chevalley complex. Then $B_1(x)=\tilde x$
\end{proposition}
\begin{proof}
The element $c_i,\tilde c_i$ are chosen to insure the identity $d\tilde x=0$:
\begin{equation}\label{E:jdghxv}
0=\sum_i(-x_ia_i\otimes b_i-a_i\otimes[x_i,b_i]+a_i\otimes[c_i,\tilde c_i])
\end{equation}
The commutator $[x_i,b_i]$ is equal to $x_ib_i+m_i$, where $m_i\in [TYM,TYM]$. We choose elements $c_i,\tilde c_i$ in such a way that $\sum_i-a_i\otimes m_i+\sum_ia_i\otimes [c_i,\tilde c_i]=0$

\end{proof}

\begin{proposition}\label{P:mxbste}
The map $\delta$ is an embedding on the image $\Imm(H_1(V,\Lambda^2(M))\subset H_2(YM,M)$. The composition $\delta^c\circ \delta:H_2(YM,M)\rightarrow H_2(YM,M)$ is a projection on $\Imm(H_1(V,\Lambda^2(M))$
\end{proposition}
\begin{proof}
The map $I_2$ is a direct image in homology induced by  abelenization.

The differential of the spectral sequence is easy to compute on the cocycle \ref{P:jhdgtwnd}. We lift $\tilde x$ to some element of $I/I^3 \otimes \Lambda^2(YM)$ and apply the differential of the later complex. The  result is
\begin{equation}
\delta (\tilde x)=\sum_i[b_i,a_i]\otimes v_i+[c_i,a_i]\otimes \tilde c_i-[\tilde c_i,a_i]\otimes c_i
\end{equation}
This element can be pushed to $H_1(YM,I^2/I^3)$ and further  to $H_1(V,M^{\otimes2})$. We get a formula:
\begin{equation}
I_2 \circ B_1 \circ \delta(\sum_i a_i\otimes b_i\wedge v_i)=\sum_i a_i\otimes b_i\wedge v_i-\sum_i b_i\otimes a_i\wedge v_i
\end{equation} 
We identify $[a_i,b_i]$(an element in $I^2/I^3$) with $a_i\otimes b_i-b_i\otimes a_i\in M^{\otimes 2}=I^2/I^3$
\end{proof}

\subsection{Restriction of $\delta$ on symmetric part of $H_2(YM,M)$}\label{S:iewgafvf}

We would like to specialize  construction of definition \ref{D:whysbxg} to  algebra $\mathfrak{h}=TYM/[TYM,[TYM,TYM]]$. The algebra $\mathfrak{h}$  is a direct sum of two linear spaces $M+\Lambda^2(M)$. The linear space $\Lambda^2(M)$ is  the center . The commutator $[.,.]:M\wedge M\rightarrow \Lambda^2(M)$ is an isomorphism.
\begin{proposition}
A linear space $D(\mathfrak{h})$ is a $\Sym(V)$-module. There is a short exact sequence of modules
\begin{equation}
0\rightarrow \Lambda^3(M) \rightarrow D(\mathfrak{h})\rightarrow \Sym^2(M)\rightarrow0
\end{equation}
\end{proposition}
\begin{proof}
A linear space $D(\mathfrak{h})$ is a quotient of $\Sym^2(M+\Lambda^2(M))=\Sym^2(M)+\Lambda^2(M)\otimes M+\Sym^2(\Lambda^2(M))$. The last summand is not present in $D(\mathfrak{h})$ because of $[a,b]\circ[c,d]=-b[a,[c,d]]=0$. A linear subspace $\Sym^2(M)$ stays intact in $D(\mathfrak{h})$. Since $[a,b]\circ c=-[a,c]\circ b$, only $\Lambda^3(M)$ part of $\Lambda^2(M)\otimes M$ survives in the quotient.

The algebra $YM$ acts on $TYM/[TYM,[TYM,TYM]]$, thus $D(\mathfrak{h})$ 	is $YM$-module. By definition commutators $[v_i,v_j]$ act trivially on every element of $a\in D(\mathfrak{h})$. It means that the action factors through abelenization of $YM$.
\end{proof}

It easy to check that the map $B_3:\Lambda^3(M)\rightarrow H_1(YM,\Lambda^2(M))\subset H_1(YM,M^{\otimes 2})$ is defined by the formula
\begin{equation}
a\wedge b\wedge c\rightarrow (a\wedge b)\otimes c+(c\wedge a)\otimes b+(b\wedge c)\otimes a
\end{equation} 
The right hand side of the last formula is Cartan-Chevalley cocycle. It is a cocycle because the action of $a,b,c\in TM$ on elements of $\Lambda^2(M)$ is trivial.

 One needs to apply some effort to convert it to $C(M^{\otimes 2})$ cocycle, which we would not do.

\begin{proposition}\label{P:lsjshcgs}
There is a commutative diagram
\begin{equation}
\mbox{
$\begin{array}{ccc}
H_2(YM,M) & \overset{\delta}{\rightarrow} & H_1(YM,M^{\otimes2})\\
\uparrow B_2& &\uparrow B_3\\
H_1(V,\Sym^2(M)) & \overset{\tilde {\delta}}{\rightarrow} & H_0(V,\Lambda^3(M))
\end{array}$
}
\end{equation}
The map $\tilde \delta$ is the boundary map corresponding to extension $D(\mathfrak{h})$
\end{proposition}
\begin{proof}
is left to the reader. It is somewhat similar to the proof of proposition \ref{P:mxbste}.
\end{proof}

\begin{proposition}\label{P:odhsgte}
Restriction of map $B_3$ on $H_0(V,\Lambda^3(M))$ has trivial kernel.
Restriction of $B_2$ on $H_1(V,\Sym^2(M))$ has kernel equal to $\Lambda^5(V)$
\end{proposition}
\begin{proof}
The kernel of mentioned map is equal to $S_2(H_2(V,M^{\otimes 2}))$. The group $H_2(V,M^{\otimes 2})$ was computed in proposition \ref{P:vsbfh}. In fact $H_2(V,M^{\otimes 2})=\Tor_{2,2}(M,M)=\Lambda^6(V)$. The second grading indicates that cocycles are  in $M_0\otimes M_0\otimes \Lambda^2(V)\cong \Lambda^2(V)^{\otimes 3}$. The tensor product $\Lambda^2(V)^{\otimes 3}$ contains one copy of $\Lambda^6(V)$ (with structure of $\mathfrak{so}_n$ representation) in $\Sym^3(\Lambda^2(V))$. The map $S_2$ transforms it into element of $\Sym^3(M)\subset M^{\otimes 3}$ and clearly misses $\Lambda^3(M)$

The second statement is proved along the same lines.
\end{proof}

\begin{proposition}
The kernel of  $\tilde {\delta}$ is $\Lambda^5(V)+V$.
\end{proposition}
\begin{proof}

Any element of $H_0(V,\Lambda^2(M))$ is a linear combination of elements of the form $F_{ij}\otimes x^{\alpha}F_{kl}-x^{\alpha}F_{kl}\otimes F_{ij}$, $\alpha$ is a multiindex. The corresponding element of $ H_1(V,\Sym^2(M))$ is equal to
\begin{equation}\label{E:mvxsdww}
\begin{split}
&\sum_s(x_sF_{ij}\otimes x^{\alpha}F_{kl} -F_{ij}\otimes x_sx^{\alpha}F_{kl})\otimes \varsigma_s\\
&\sum_s(x^{\alpha}F_{kl}\otimes x_sF_{ij}- x_sx^{\alpha}F_{kl}\otimes F_{ij})\otimes \varsigma_s
\end{split}
\end{equation}
From commutativity of \ref{P:lsjshcgs} and proposition \ref{P:odhsgte}, linear space $\Lambda^5(V)$ must be in a kernel of $\tilde \delta$. The check of $\tilde\delta a=0$, where $a$ is defined in \ref{E:shsc} is left to the reader as an exercise. The cocycles \ref{E:chdcs} are skewsymmetric and will be ignored in present discussion. Thus studying the kernel of $\tilde \delta$ we can safely restrict $\tilde \delta$ on span of elements \ref{E:mvxsdww}.
According to corollary \ref{E:lfgjdw} there is a surjective map $H_0(V,M^{\otimes 2})\rightarrow H_1(V,M^{\otimes 2})/\Lambda^5(V)+\Lambda^3(V)+V$. It splits into a direct sum $H_0(V,\Sym^2(M))\rightarrow H_1(V,\Lambda^2(M))/\Lambda^3(V)$ and  $H_0(V,\Lambda^2(M))\rightarrow H_1(V,\Sym^2(M))/\Lambda^5(V)+V$.

We need to set some more notations. We know that $U(TYM)\cong T(M)$. The identification is not quite canonical. There are a filtrations $F^i$ on $U(TYM)$ and on $T(M)$; the isomorphism is compatible with filtrations and independent of any choices on $Gr$'s.

The isomorphism enables us to transfer the action of $YM$ from $U(TYM)$ to $T(M)$. This operation is defined up  to mentioned above ambiguity.

The information we already possess enables us to write the action $\rho(v_i)$ of generators $v_i$ of $YM$ on algebraic generators  $m\in M\subset T(M)$:
\begin{equation}\label{E:hsgxcv}
\rho(v_i)m=x_im+\psi_i^2(m)+\psi_i^3(m)+\dots
\end{equation}
In this formula $x_im$ is the action of generator $x_i\in \Sym(V)$ on element of $\Sym(V)$-module $M$, $\psi^k_i:M\rightarrow Free^k(M)\subset M^{\otimes k}, k=2,\dots$. 

The algebra $TYM\subset YM$ acts on $T(M)$ by inner derivations. Denote $F_{ij}$-$\Sym(V)$ generators of $M$. Then 
\begin{equation}
[\rho(v_i),\rho(v_j)]m=F_{ij}m-mF_{ij}
\end{equation}
This identity gives some restrictions on maps $\psi^k_i$.

In terms of  $\psi_i=\psi^2_i$ it is easy to write the differential $\tilde \delta$.
\begin{equation}
\begin{split}
&\tilde \delta x=\\
&\sum_s (\psi_s(x_sF_{ij})\otimes x^{\alpha}F_{kl}+   x_sF_{ij}\otimes\psi_s( x^{\alpha}F_{kl}))  \quad\quad (a_1+b_1) \\
&\sum_s(-\psi_s(F_{ij})\otimes x_sx^{\alpha}F_{kl}-  F_{ij}\otimes\psi_s( x_sx^{\alpha}F_{kl}))  \quad\quad (a_2+b_2)\\
&\sum_s(\psi_s(x^{\alpha}F_{kl})\otimes x_sF_{ij}+   x^{\alpha}F_{kl}\otimes \psi_s(x_sF_{ij})) \quad\quad (a_3+b_3)\\
&\sum_s(- \psi_s(x_sx^{\alpha}F_{kl})\otimes F_{ij}-  x_sx^{\alpha}F_{kl}\otimes \psi_s(F_{ij}))\quad\quad (a_4+b_4)
\end{split}
\end{equation}
The element $\tilde \delta x$ belongs to the group $H_0(V,\Lambda^3(M))$, which gives some freedom in algebraic manipulations. In particular 
\begin{equation}\label{E:zxcsea}
a_1+a_2\sim \left( \sum_s \psi_s(x_sF_{ij})+x_s\psi_s(F_{ij})\right) \otimes x^{\alpha}F_{kl}
\end{equation}
 There are similar formulas for $b_1+b_2$, $b_3+b_4$, $a_3+a_4$. 
To simplify this formula we need more information about $\psi_s$.

Suppose $m\in M$. Let us write a formula for $$\sum_s\rho(v_s)\rho(v_s)m\overset{def}{=}\Delta(m)$$  We have
\begin{equation}\label{E:orjhcd}
\Delta(m)=\sum_sx_s\psi_s(m)+\psi_s(x_sm)+c(m), \quad c(m)\in F^3
\end{equation}
The same form has left tensor factor in \ref{E:zxcsea}.
Let is analyze how the formula \ref{E:orjhcd} depends on the identification $U(TUM)$ and $T(M)$. The identification is determined by embedding $\mu:M\rightarrow U(TYM)$. It must satisfy the condition: the composition $M\overset{\mu}{\rightarrow}F^1/M^2\rightarrow M$ must be identity map. It is easy to see that all such embedding is a homogeneous space under the group $G$ of automorphisms of $U(TYM)$ which preserves the filtration $F^i$ and whose  action on $Gr(U(TYM))$ is trivial. If we choose a reference point $\mu$(or isomorphism $U(TUM)$ and $T(M)$) we can identify the homogeneous space with the group $G$. An element $\beta$ is uniquely determined by its values on generating space $M$:
\begin{equation}
\beta(m)=m+\alpha^2(m)+\alpha^3(m)+\dots,\quad \alpha^i(m)\in M^{\otimes i}
\end{equation}
In the following we will neglect elements in $\bigoplus_{j\geq 3}M^{\otimes j}$, therefor to simplify notations we set $\alpha^2=\alpha$.
The inverse automorphism has a form $\beta^{-1}(m)=m-\alpha(m)+\dots$. Thus $\alpha\Delta\alpha^{-1}(m)=\sum_s(x_s\psi_s(m)+\psi_s(x_sm)-x_s^2\alpha(m))+\dots$. From the last formula we see that the leading term of $\Delta$ is defined up to addition of $\sum_sx_s^2(m_i\otimes m'_i)$. This will not change the element $\delta x$ in $H_0(V,\Lambda^3(M))$, because for example in $a_1+a_3$-term we have $\sum_sx_s^2\alpha(F_{ij})\otimes x^{\alpha}F_{kl}\sim \sum_s\alpha(F_{ij})\otimes x^2_sx^{\alpha}F_{kl}=0$

To make our computations more tractable we need to make further simplification: we will map $H_0(V,\Lambda^3(M))$ to $\Lambda^2(V)\otimes H_0(V,\Lambda^2(M))$.

Indeed there is a map of $\Sym(V)$-modules $p:\Lambda^3(M)\rightarrow \Lambda^2(V)\otimes\Lambda^2(M)$, with trivial action on $\Lambda^2(V)$-factor. The homomorphism is defined as:
\begin{equation}
\begin{split}
&p(aF_{ij}\wedge bF_{kl}\wedge cF_{st})=\\
&=a(0)\hat F_{ij}\wedge bF_{kl}\wedge cF_{st}+\\
&+aF_{ij}\wedge b(0)\hat F_{kl}\wedge cF_{st}+\\
&+aF_{ij}\wedge bF_{kl}\wedge c(0)\hat F_{st}
\end{split}
\end{equation}
In the last formula $\hat F_{ij}$ should be considered as an element of $\Lambda^2(V)$. It is easy to check that this is an  unambiguously defined  homomorphism of $\Sym(V)$-modules. This homomorphism produces a map of homologies $p:H_0(V,\Lambda^3(M))\rightarrow \Lambda^2(V)\otimes H_0(V,\Lambda^2(M))$

As we know (see \cite{MSch2}) the components $M_i$ of graded $\Sym(V)$-module $M$ has  the following geometric interpretation: these are sections of $\mathfrak{so}_n$-homogeneous vector bundle $T(i)$-${\cal O}(i)$-twisted tangent bundle  on a nonsingular quadric $Q$. We identify $\Lambda^2(V)$ with Lie algebra  $\mathfrak{so}_n$. The operator acting on $M$ that corresponds to $\hat F_{ij}\in \mathfrak{so}_n$ we denote by $\xi_{ij}$.
\begin{lemma}\label{L:tyebcj}

The following formula holds:
\begin{equation}
p(\tilde\delta x)=4\sum_{s<t}\hat F_{st}\otimes \xi_{st}x
\end{equation}
\end{lemma}

\begin{proof}

A simplification pointed in \ref{E:zxcsea} leads  to 
\begin{equation}
\begin{split}
&\tilde \delta x=\\
&\Delta F_{ij}\otimes x^{\alpha}F_{kl}-\Delta x^{\alpha}F_{kl}\otimes F_{ij}- \\
&-F_{ij}\otimes\Delta x^{\alpha} F_{kl}+x^{\alpha}F_{kl}\otimes\Delta F_{ij}
\end{split}
\end{equation}

Therefor it is crucial to compute $\Delta x^{\alpha}F_{ij}$. Write $\Delta x^{\alpha}F_{ij}=\Delta x_{i_1}\dots x_{i_k}F_{ij}, |\alpha|=k$. We lift it to 
\begin{equation}\label{E:owbxrw}
\Delta x^{\alpha}F_{ij} \Rightarrow \sum_s[x_s,[v_s[v_{i_1},\dots[v_{i_k}, F_{ij}]\dots]
\end{equation}
 We replace commutators $[v_i,y],y\in T(M)$ by $x_iy+\psi_iy$ and drop terms which belong to $F^3$. We see that 
\begin{equation}
\sum_s[v_s,[v_s[v_{i_1},\dots[v_{i_k}, F_{ij}]\dots]\sim \Delta(x^{\alpha}F_{ij})+\sum_{t=1}^k\sum_s x^2_sx_{i_1}\dots\psi_{i_t}(\dots x_{ik}F_{ij})
\end{equation}
which justifies the choice of the lifting \ref{E:owbxrw}. On the other hand \ref{E:owbxrw} is equal to 
\begin{equation}\label{E:kfhjdg}
\begin{split}
&\sum_{t=1}^k\sum_s([v_s,[\dots [v_{i_{t-1}},[F_{si_t}[v_{i_{t+1}},[\dots F_{ij}]\dots])+\sum_s[v_s,[v_1,[\dots,[v_s,F_{i,j}]\dots]\sim\\
&\sim 2\sum_{t=1}^k\sum_s x_sx_{i_1}\dots x_{i_{t-1}}([F_{si_t},x_{i_{t_1}}\dots F_{ij}])+2\sum_{s}x^{\alpha}[F_{si},F_{sj}]
\end{split}
\end{equation}
We use the formula \ref{E:kfhjdg} to compute $p(\delta x)$

Making manipulations similar  to \ref{E:zxcsea} we have
\begin{equation}
\begin{split}
&p(a_1+a_2)=2\hat F_{si}\otimes F_{sj}\wedge x^{\alpha}F_{kl}-2\hat Fi_{sj}\otimes F_{si}\wedge x^{\alpha}F_{kl}\\
&p(a_3+a_4)=2\hat F_{sk}\otimes F_{sj}\wedge x^{\alpha}F_{sl}-2\hat F_{sl}\otimes F_{ij}\wedge x^{\alpha}F_{sk}\\
&p(b_1+b_2)=2\hat F_{sk}\otimes F_{ij}\wedge x^{\alpha}F_{sl}-2\hat F_{sl}\otimes F_{ij}\wedge x^{\alpha}F_{sk}+2\hat F_{st}\otimes F_{ij}\wedge (\xi_{st}x^{\alpha})F_{kl}\\
&p(b_3+b_4)=2\hat F_{st}\otimes F_{ij}\wedge (\xi_{st}x^{\alpha})F_{kl}+2\hat F_{si}\otimes F_{sj}\wedge x^{\alpha}F_{kl}-2\hat F_{sj}\otimes F_{si}\wedge x^{\alpha}F_{kl}
\end{split}
\end{equation}
From this we conclude that 
\begin{equation}
p(a_1+a_2+a_2+a_4+b_1+b_3+b_2+b_4)=4\sum_{s<t}\hat F_{st}\otimes \xi_{st}(F_{ij}\wedge x^{\alpha}F_{kl})
\end{equation}

\end{proof}

The action of $\mathfrak{so}_n$ on $Tor_0(M,M)=H_0(V,M^{\otimes 2})$ according to proposition \ref{E:vedvxjkff} has no trivial subrepresentations.
From lemma \ref{L:tyebcj} the composition $p\tilde \delta$ is injective, thus $\tilde\delta$ is injective
\end{proof}

\section{Computation of $H^i(YM,U(YM)), \quad i=0,1$}\label{S:oquebnx}
The  goal of this section is to prove proposition \ref{P:dhfggn}
\begin{lemma}
\begin{align}
&H^1(YM,U(YM))\supset \mathbb{C}+V+\Lambda^2(V)\label{P:affsgsdgd}\\
&H^1(YM,YM)\supset \mathbb{C}+\Lambda^2(V)\label{P:afgfsgsdgd}
\end{align}
\end{lemma}
\begin{proof}

Let us make some simple observations. Algebra $YM$ is graded, hence we have one differentiation for free: differentiation corresponding to grading. This explains $\mathbb{C}$ in \ref{P:affsgsdgd}

The linear space of relation \ref{E:jsjhsgtw} of $YM$ is invariant  with respect to the action of $\mathfrak{so}_{n}$. The adjoint representation of  $\mathfrak{so}_{n}$ is equal to $\Lambda^2(N)$. It explains the last summand in \ref{P:affsgsdgd}.

Derivation $D_s(v_t)=\delta_{st}$ is compatible with relation \ref{E:jsjhsgtw}. The linear space generated by  $D_s$ is isomorphic to $V$ as  $\mathfrak{so}_{n}$ representation. It is a derivation of $U(YM)$. It explains why $V$ factor is not present in \ref{P:afgfsgsdgd}.

\end{proof}

Let $\mathfrak{h}$ be a Lie algebra of symmetries of a Lie algebra $\mathfrak{g}$.
Due to Poincare-Birkhoff-Witt theorem there is an isomorphism of $\mathfrak{h}$ modules under adjoint action 
\begin{equation}\label{E:sdxjv}
U(\mathfrak{g})=\Sym(\mathfrak{g})
\end{equation}
Denote the symmetric product in $\Sym(\mathfrak{h})$ by $\circ$.

To compute cohomology with coefficients in $U(YM)$, we can safely replace last module by $\Sym(YM)$.

Consider filtration $F^i$ of $\Sym(YM)$ defined by powers of an ideal generated by the linear space $TYM$. It defines a filtration on $\Sym^n(YM)$, for which we use the same notation.  It adjoint quotients are equal $F^i/F^{i+1}=\Sym^{n-i}(V)\otimes \Sym^i(TYM)$. 

The filtration gives rise to a spectral sequence $E^{ij}_r$. There is an easy established formula for  $E^{ij}_1$:
\begin{equation}
E^{ij}_1=H^{i+j}(YM,\Sym^{n-i}(V)\otimes\Sym^i(TYM)
\end{equation}
The action of $YM$ on $\Sym^{n-i}(V)$-factor is trivial.

\begin{proof}{\bf of proposition \ref{P:dhfggn}}
We use mentioned spectral sequence for computation $H^1(YM,\Sym(YM))$. Our prime interest will be the fragment $E^{ij}_r$, $i+j=0,1$, because only it contribute to $H^k(YM,\Sym(YM)), k=0,1$.

Computations of the previous section provide us with necessary information about  $E^{ij}_1$ . It is reformulated in next two lemmas in a convenient form.
\begin{lemma}\label{L:jdbc}
$H^1(YM,\Sym^i(TYM))=0$ $i\geq 2$, $H^1(YM,\Sym^i(TYM))=V$ $i=0,1$ and $H^0(YM,\Sym^i(TYM))=0, i\geq 1$
\end{lemma}
\begin{proof}
Apply  isomorphism \ref{E:sdxjv} to $\mathfrak{g}=TYM$. Set $F_{ij}=[v_i,v_j]$ in algebra $YM$. The  cocycles $\sum_s F_{ks}\otimes A^{*s}$, $k=1,\dots,n$ of $C(U(TYM))$ (see \ref{nnasf}), which span $H^1(YM,U(TYM))$ according to proposition \ref{P:ududufgqex}, belong to $H^1(YM,\Sym^1(TYM))$. The later  is a direct summand of $\Sym(TYM)$. From this we conclude that $H^1(YM,\Sym^i(TYM))=0$ $i\geq 2$. $i=0$-case was treated in details in \cite{MSch2}. In particular it implies that that $E^{i,1-i}_1=0$ for $i\geq 2$.

The statement about zeroth cohomology follows from proposition \ref{P:ududufgqex}.
\end{proof}

\begin{lemma}
$H^1(YM,\Sym^i(V)\otimes\Sym^k(TYM))$ is equal to $\Sym^i(V)\otimes V$ if $k=0,1$ and $0$ if $k>1$ .

We also have trivial isomorphisms $H^0(YM, \Sym(V))=\Sym(V)$
\end{lemma}
\begin{proof}
Easily follows from \ref{L:jdbc}
\end{proof}

We would like to compute differential $d:E^{i,j}_1\rightarrow E^{i+1,j}_1$. Information about this differential is contained in the following two extensions (for the range $i,j$ we are interested in):
\begin{equation}
TYM\otimes \Sym^{n-1}(V)\rightarrow E_n \rightarrow \Sym^n(V), E_n=F^0/F^{2}
\end{equation}
and
\begin{equation}\label{E:xhshbx}
\Sym^2(TYM)\otimes \Sym^{n-2}(V)\rightarrow G_n\rightarrow TYM\otimes \Sym^{n-1}(V), G_n=F^1/F^{2}
\end{equation}

For $n=2$ set $G=G_2$, $E=E_2$. Consider a boundary differential $\delta:H^1(YM, V \otimes TYM)\rightarrow H^1(YM,\Sym^2(TYM))$ in a long exact sequence of cohomology of $YM$ associated with \ref{E:xhshbx}. By definition of $d_1$- the differential of $E^{ij}_1$, it appropriate component is equal to $\delta$.
\begin{lemma}\label{L:fjiudj}
The image of $\delta$ is equal to $\Lambda^2(V)$
\end{lemma}
\begin{proof}
There is also a boundary map $\tilde\delta:H^0(YM,\Sym^2(V))=E^{00}_1\rightarrow E^{10}_1=H^1(YM,V \otimes TYM)$ associated with extension $E$. It is also a component of $d_1$.

We already know that $\tilde \delta:H^0(YM,V)\rightarrow H^1(YM,TYM)$ is an isomorphism, hence $\delta:H^0(YM,\Sym^2(V))\rightarrow H^1(YM,V \otimes TYM)$ is an embedding.

An equation $\delta\tilde\delta=0$ is a corollary of $d_1^2=0$. From this  $\delta \Sym^2(V)=0$ trivially follows.
Thus the map $\delta$ factors through $\Lambda^2(V)$. 

In  realization $H(YM,G)$ defined by  $C(G)$ ( see formula \ref{nnasf}) an identity folds:
\begin{equation}
\begin{split}
&\delta \left( \sum_s(v_1\circ [v_2,v_s]-v_2\circ [v_1,v_s])\otimes A^{*s}\right)=\\
&\sum_{st}[v_1.F_{st}]\circ [v_2,v_s]-[v_2.F_{st}]\circ [v_1,v_s]\in C^2(V \otimes TYM)
\end{split}
\end{equation}

This is nontrivial $C(V \otimes TYM)$ cocycle proved by degree counting arguments.

\end{proof}

We would like to generalize this statement. As in case $n=2$ we prove that $\delta:H^0(YM,\Sym^n(V))\rightarrow H^1(YM, \Sym^{n-1}(V) \otimes TYM$ is an embedding. Due to identifications $H^0(YM,\Sym^n(V))=\Sym^n(V)$ and $H^1(YM,\Sym^{n-1}(V) \otimes TYM )=\Sym^{n-1}(V)\otimes V$, we can interpret  the boundary map $\delta=d_1$ as ah de Rham differential $d_{dR}$ mapping polynomial functions to polynomial one-forms.

The differential $d_1$ satisfies
\begin{equation}
d_1(ab)=d_1(a)b+ad_1(b), a\in H^0(YM,\Sym^{m}(V)), b\in H^1(YM,\Sym^{k}(V) \otimes TYM)
\end{equation}
for suitable $m,k$. This is a standard property of differential in a spectral sequence produced by multiplicative filtration.


\begin{lemma}
Suppose we have a map $d_1 :\Sym(V)\otimes \Lambda^1(V)\rightarrow \Sym(V)\otimes \Lambda^2(V)$, such that $d_1(ab)=d_{dR}ab+ad_1b$, $a\in \Sym(V)$, $d_1 d_{dR}=0$ and $d_1(x_id_{dR}x_j)=d_{dR}x_i\wedge d_{dR}x_j$, then $d_1=d_{dR}$. 
\end{lemma}
\begin{proof}
Left to the reader.
\end{proof}

\begin{corollary}
The complex $H^0(YM, \Sym(V))\overset{d_1}{\rightarrow} H^1(YM, \Sym(V)\otimes TYM)\overset{d_1}{\rightarrow} H^2(YM, \Sym(V)\otimes \Sym^2(TYM))$ is acyclic in the middle term. The zeroth cohomology is equal to $\mathbb{C}$.
\end{corollary}

We can interpret this corollary as no subquotients of $E^{1,0}_1$ give contributions to $H^1(YM,\Sym^n(YM))$. 
Only  $H^0(YM,\mathbb{C})$ gives contribution to  $H^0(YM,U(YM))$. This observation proves the 
\begin{corollary}
$H^0(YM,U(YM))=\mathbb{C}$
\end{corollary}

 An alternative spot in the spectral sequence that can potentially contribute to $H^1(YM,\Sym^n(YM)$ is $E^{0,1}_1=H^1(YM,\Sym^n(V))$. We proceed with analysis of this remaining case.

\begin{lemma}
The boundary map $\delta:H^1(YM, V)\rightarrow H^2(YM, TYM)$ associated with extension $E_1$ has image equal to $\Sym^2(V)/\mathbb{C}\delta_{ij}$. $\delta_{ij}$ stands for a symmetric bilinear form on $V$ which is invariant with respect to $\mathfrak{so}_n$
\end{lemma}
\begin{proof}
Direct inspection of the complex $C(E_1)$ similar to the carried out in the proof of \ref{L:fjiudj}.
\end{proof}

Due to multiplicatively of $d_1$ one can decompose the boundary differential associated with $E_n$ into  composition 
\begin{equation}\label{E:jsbcz}
\begin{split}
&H^1(YM, \Sym^n(V))=\Sym^n(V)\otimes V\overset{d_{dR}\otimes 1}{\rightarrow} \Sym^{n-1}(V)\otimes V^{\otimes 2}\overset{1\otimes p}{\rightarrow}\\
&\rightarrow \Sym^{n-1}(V)\otimes \Sym^2(V)/\mathbb{C}\delta_{ij}\subset H^2(YM, \Sym^{n-1}(V)\otimes TYM)
\end{split}
\end{equation}
The map $p$ is projection $p:V^{\otimes 2}\rightarrow\Sym^2(V)/\mathbb{C}\delta_{ij}$.
$d_1$ has the following geometric interpretation: the linear space $\Sym(V)\otimes V$ can be identified with a space of polynomial vector fields on $\mathbb{C}^n$, $\Sym(V)\otimes \Sym^2(V)/\mathbb{C}\delta_{ij}$-space of polynomial traceless (with respect to the standard metric $\delta_{ij}$) symmetric two-tensors. The value of composed map $d_1(\xi)$ is a traceless part of the Lie derivative- $L_{\xi}\delta_{ij}$.

\begin{lemma}
The kernel of the map $d_1$ in \ref{E:jsbcz} is a Lie algebra of  conformal Lie group.
\end{lemma}
\begin{proof}
This can be considered as a definition . If $dimV\geq 3$ then the following vector fields form a basis in such space:

\begin{equation}
\begin{split}
&\pr{x_i} \mbox{-shifts}\\
&x_i\pr{x_j}-x_j\pr{x_i} \mbox{- rotations}\\
&\sum_{i=1}^n\left(x_kx_i\pr{x_i}-1/2x_i^2\pr{x_k}\right)=\sum_s a_{ks}\pr{x_s}\mbox{ -conformal vector fields}\\
&\sum_{i=1}^nx_i\pr{x_i}\mbox{ -dilation}
\end{split}
\end{equation}
\end{proof}

An interpretation of this lemma in terms of our spectral sequence is: only Lie algebra of conformal group equal to $E^{0,1}_2$has a chance to contribute to $H^1(YM,\Sym(YM))$. We already know that translations, rotations and dilations are symmetries of $YM$; they represent nontrivial cocycles in  $H^1(YM,\Sym(YM))$ and survive to the limiting term of the spectral sequence. It is not so clear for conformal vector fields. In the remaining part of the proof we shall explore this. 

We associate with a conformal vector field $\sum_sa_{ks}\pr{x_s}$ a one-cocycle $\sum_sa_{ks}\otimes A^{*s} \in C(\Sym^2(V))$. The fact that $d_1(a)=0$ means that the cocycle can be lifted to a cocycle with values in $E=F^0/F^2$. We would like to know if it is possible to lift it to cocycle with values in $\Sym^2(YM)$. To do that it is suffice to compute a boundary operator associated with extension $\Sym^2(TYM)\rightarrow \Sym^2(YM)\rightarrow E$.
\begin{lemma}
The conformal vector fields could not be lifted to one-cocycles in $C^1(\Sym^2(YM))$
\end{lemma}
\begin{proof}

The boundary operator maps $H^1(YM,E)\rightarrow H^2(YM,\Sym^2(TYM))$ . The result of computation is

\begin{equation}
\delta(a_{ks}A^{*s})=4[v_s,v_k]\circ [v_s,v_t]\otimes A_t-[v_s,v_i]\circ [v_s,v_i]\otimes A_k\in C^2(\Sym^2(TYM))
\end{equation}
By grading counting this is nontrivial cocycle in $C(\Sym^2(TYM))$. Hence conformal vector fields could not be lifted to one-cocycles of $YM$ with values in $\Sym^2(YM)$. 
\end{proof}

This computation is equivalent to computation of components $d_2:E^{0,1}_2\rightarrow E^{2,0}_2$ of differential $d_2$ in the second term $E^{ij}_2$ of the spectral sequence. The computation tells us that conformal vector field does not survive to the second term.

As a result of our computation of $H^1(YM,\Sym(YM))$ we are left with cocycles with values in $\Sym^0(YM), \Sym^1(YM)$. 

The first group is generated by shifts, the second group is generated by rotation and dilation.
\end{proof}

\section{Appendix}\label{S:oqowe}
\subsection{Generating functions of algebra $YM$}\label{SS:sjshd}

The algebra $YM$ is graded, so it makes sense to talk about Poincare series $\sum dimU(YM)_it^i$. The sum of this series can be computed from acyclicity of resolution (see \cite{MSch2}) :
\begin{equation}
U(YM)_i\rightarrow U(YM)_{i+2}\otimes V\rightarrow U(YM)_{i+6}\otimes V \rightarrow U(YM)_{i+8}
\end{equation}
 We have the following formula:
\begin{equation}
\sum dimU(YM)_it^i=\frac{1}{1-dimVt^2+dimVt^6-t^8}
\end{equation} 

\subsection{Generating functions of module $M$}\label{S:bhshhdshd}
Denote $M(t)=\sum_{i\geq 0}dimM_it^i$.
\begin{proposition}\label{P:fasdhfjh}
$1-t^2M(t)=\frac{1-dimVt+dimVt^3-t^4}{(1-t)^{dimV}}$
\end{proposition}
\begin{proof}
The complex $C^i(\Sym(V))$ splits into a direct sum of homogeneous components: $C^i(\Sym(V))=\bigoplus_{j}C_j^i(\Sym(V))$. The Euler characteristic is $$\chi(t)=\sum_{ij}(-1)^{i+j}dimC_j^i(\Sym(V))t^j$$ equal to $-(1-t^2M(t))$ because of the cohomology computations (proposition \ref{P:cxbvcvg}). By myltiplicativity $\chi(t)=-\Sym(V)(t)(1-dimVt+dimVt^3-t^4)$, It is well known that $\Sym(V)(t)=\frac{1}{(1-t)^{dimV}}$.
\end{proof}

\subsection{Cyclic homology of multigraded free algebras.}
The formulas presented in this section are not new and should be well known to specialists. The author decided to present them here because of inability to find a reference.

Denote by $A$ positively multigraded vector space $A=\bigoplus A_{\alpha}$, where $\alpha=(i_1,\dots ,i_k)$-s a multiindex and $|\alpha|=\sum_{j=1}^k i_j$. Then $A(z)=A(z_1,\dots ,z_k)=\sum_{\alpha}dim(A_{\alpha})z^{\alpha}$ is Poincare series.
We say that positive integer $s$ divides multiindex $\alpha=(i_1,\dots ,i_k)$ (denote $s|\alpha$) if $s$ divides all $i_j$. Denote $GCD(\alpha)$ the greatest common denominator of $i_j$.

Denote $Cyc(V)=T(V)/[T(V),T(V)]$ a linear space of cyclic words over an alphabet, formed by some basis of $V$, $\overline{Cyc}(V)=Cyc(V)/\mathbb{C}$.
\begin{proposition}\label{P:kjcjsh}
Suppose $V$ is a positively multigraded vector space .
Let $T(V)$ be a free algebra on $V$.
The reduced cyclic homology $\underline{HC}_i(T(V))$ are equal to zero for $i>0$ and 
\begin{equation}\label{LLom1}
\begin{split}
&\underline{HC}_0(T(V))(z_1,\dots ,z_k)=\overline{Cyc}(V)(z_1,\dots ,z_k)=\\
&=-\sum _{t\geq 0}\frac {\psi (t)}{t} ln (1-V((-1)^{t+1}z_1 ^t,\dots ,(-1)^{t+1}z_k^t))
\end{split}
\end{equation}
where $\psi (t)$ is the Euler psi-function (i.e. $\psi (t)$ counts the number of integers which are less then $t$ and co-prime with $t$).
\end{proposition}

\begin{proof}
The free algebra $T(V)$ is a universal enveloping of a free Lie algebra $Free(V)$.
As a $Free(V)$ module under adjoint action 
, we have an isomorphism $T(V)=\oplus _{i\geq0}\Sym^i Free(V)$.

To compute Hochschild homology of $Free(V)$ with coefficients in module $M$ we can follow \cite{Loday} and use a complex 
\begin{equation}
K=(V\otimes M\overset {d}{\rightarrow}M),
\end{equation}
 where the differential is defined as $a\otimes m\rightarrow am$.
Let $M=\Sym Free(V)$. We have an isomorphism $H^0(K(\Sym (Free(V))))=HH_0(T(V))=HC_0(T(V))$ with a refined decomposition $$HC_0(T(V))=\bigoplus_{k>0}HC_0^k(T(V))=\bigoplus_{k>0}H^0(K(\Sym ^k(Free(V))))$$
According to \cite{Loday} $HH_i(T(V))=0\quad i\geq 2$. For small values of $i$ we have Connes exact sequence, which for graded algebras splits into isomorphism $$HH_1(T(V))=\overline{HC}_0(T(V))$$ and $$HH_0(T(V))=HC_0(T(V)$$. Denote $HC^k_0=H_0(K(\Sym ^k(Free(V))$, then (see \cite{Loday})  $H_1(K(\Sym ^kFree(V))=\overline{HC}_0^{k+1}$. 
We can write the following equation for generating functions:
\begin{align}
&\sum _{i\geq 0}\chi (K(\Sym ^i))(z_1,\dots ,z_n)t^i=[1-V(z_1,\dots ,z_n)](\Sym (t,z_1,\dots ,z_n)=\notag \\
&=\sum t^k(HC_0^k(z_1,\dots ,z_n)-HC_0^{k+1}(z_1,\dots ,z_n)=\notag \\
&=\sum ^{\infty}_{k=0}t^kHC_0^k(z_1,\dots ,z_n)-\frac {1}{t}\sum ^{\infty}_{k=1}t^kHC_0^k(z_1,\dots ,z_n)=\notag \\
&=(1-\frac {1}{t})\sum ^{\infty}_{k=1}t^kHC^k(z_1,\dots ,z_n)+1\notag \\
&\mbox{ we infer }\notag \\
&HC(t,z_1,\dots ,z_n)=\frac {t((\Sym (t,z_1,\dots ,z_n)(1-V(z_1,\dots ,z_n))-1)}{t-1}\notag 
\end{align}
To compute $HC(1,z_1,\dots ,z_n)$ we need to let $t\rightarrow 1$.
To do so we can use l'Hopitale rule 
\begin{equation}\label{E:qdadxctd}
HC(1,z_1,\dots ,z_n)=\frac {\partial \Sym (t,z_1,\dots ,z_n)}{\partial t}|_{t=1}(1-V(z_1,\dots ,z_n)).
\end{equation}

For any positively multigraded vector space $A$  the generating function $\Sym(A)(t,z_1,\dots ,z_n)=\sum ^{\infty}_{i=0}\Sym ^i (A)(z_1,\dots ,z_n)t^i$ can be computed by the formula 
$
\Sym(A)(t,z_1,\dots ,z_n)=\frac{\prod_{|\alpha|odd}(1+tz^{\alpha})^{A_{\alpha}}}{\prod_{|\alpha |even}(1-tz^{\alpha})^{A_{\alpha}}}
$
So 
\begin{equation}\label{E:ajadjhc}
\Sym(A)(t,-z_1,\dots ,-z_n)=\frac {\prod _{|\alpha|odd}(1-tz^{\alpha})^{A_{\alpha}}}{\prod_{|\alpha |even}(1-tz^{\alpha})^{A_{\alpha}}}
\end{equation}
Using logarithmic derivative we have 
\begin{equation}
\begin{split}
&\frac {\partial}{\partial t}ln \Sym(A) (t,-z_1,\dots ,-z_n)=\sum_{|\alpha|odd}\frac {-tz^{\alpha}A_{\alpha}}{1-tz^{\alpha}}+\sum _{|\alpha |even}\frac {tz^{\alpha}A_{\alpha}}{1-tz^{\alpha}}=\\
&=\sum _{\alpha}\frac {(-1)^{|\alpha|}z^{\alpha}tA_{\alpha}}{1-tz^{\alpha}}.
\end{split}
\end{equation}
Then $$\frac {\partial}{\partial t}\Sym(A) (t,-z_1,\dots ,-z_n)|_{t=1}=\frac {\sum(-1)^{|\alpha|}z^{\alpha}A_{\alpha}}{1-z^{\alpha}}\Sym(A)(1,-z_1,\dots ,-z_n)$$

We would like to specialize this construction to the case $A=Free(V)$.
We have  $\Sym(Free(V)) (1,-z_1,\dots ,-z_n)=\frac {1}{1-V(-z_1,\dots ,-z_n)}$ because $\Sym (Free(V))=T(V)$.
Denote $f_{\alpha}=dim(Free(V)_{\alpha})$. By \ref{E:qdadxctd}

\begin{equation}\label{E:hssvxgg}
\begin{split}
&HC(1,-z_1,\dots ,-z_n)=\sum _{\alpha}\frac {(-1)^{|\alpha|}z^{\alpha}f_{\alpha}}{1-z^{\alpha}}=\sum_{\alpha} (-1)^{|\alpha|}f_{\alpha}\sum_{n\geq 1}z^{\alpha n}=\\
&=\sum_{\beta} z^{\beta}\sum _{n|\beta}(-1)^{\frac {|\beta|}{n}}f_{\frac {\beta}{n}}.
\end{split}
\end{equation}
The following manipulations follow from \ref{E:ajadjhc}:
\begin{align}
&ln \frac {1}{1-V(-z_1,\dots ,-z_n)}=-ln (1-V(-z_1,\dots ,-z_n))=\\
&=\sum _{|\alpha|odd}f_{\alpha}ln (1-z^{\alpha})-\sum _{|\alpha|even}f_{\alpha}ln (1-z^{\alpha})=\sum _{\alpha}(-1)^{|\alpha|}f_{\alpha}\sum_{n>0}\frac {z^{\alpha n}}{n}=\\
&=\sum z^{\beta}\sum _{k|\beta}(-1)^{\frac {|\beta|}{k}}\frac{f_{\frac {\beta}{k}}}{k}.
\end{align}

We need to remind some properties of $\psi (t)$.
There is a main identity 
\begin{equation}
\sum _{nk=l}\frac {\psi (k)}{nk}=1
\end{equation}
Hence

\begin{equation}
\begin{split}
&-\sum^{\infty}_{n=0}ln[ 1-V(-z^n,\dots ,-z^n)]\frac {\psi (n)}{n}=\sum _{\alpha}(-1)^{|\alpha|}f_{\alpha}\sum _{k,n>0}\frac {z^{\alpha nk}}{nk}\psi (k)=\\
&=\sum z^{\beta}\sum _{\mbox{\scriptsize{$\begin{array}{l}\beta = \beta_0l\\ GCD(\beta_0)=1\\ nkm=l\end{array}$}}}(-1)^{|\beta _0|m}f_{\beta _0 m}\sum _{nk=\frac {l}{m}}\frac {1}{n}\frac {\psi (k)}{k}=\\
&=\sum z^{\beta}\sum _{\mbox{\scriptsize{$\begin{array}{l} \beta =\beta_0l\\ m|l\end{array}$}}}(-1)^{|\beta _0|m}f_{\beta_0m}=HC(1,-z_1,\dots ,-z_n)
\end{split}
\end{equation}
The last identity is due to \ref{E:hssvxgg}
\end{proof}

\subsection{Some applications}
In the following text we shall discuss some applications of the formula \ref{LLom1}.

Suppose $T(V)$ is a free algebra on a bigraded vector space $V$.
Assume that $T(V)$ is equipped with a differential of degree $-1$ with respect to the first grading and homogeneous with respect to the second.
The differential induces a differential on the cyclic homology of $T(V)$.
Using formula (\ref{LLom1}) we can easily compute Euler characteristic of $HC(T(V),d)$.
\begin{proposition}
\begin{equation}\label{E:iushbcgh}
\chi \overline{HC}(T(V),d)(z)=-\sum_{n>0}ln(1-V(-1,(-1)^{n+1}z^n))\frac {\psi (n)}{n} 
\end{equation}
\end{proposition}
\begin{proof}
Is an obvious application of proposition \ref{P:kjcjsh}
\end{proof}

\begin{proposition}
Suppose $A$ is a positively graded algebra.
We have $\chi \overline{HC}(A)=-\sum ln (\frac {1}{A((-1)^{n+1}z^n)})^{\frac {\psi (n)}{n}}=\sum ln A((-1)^{n+1}z^n)\frac {\psi (n)}{n}$.
\end{proposition}

\begin{proof}
%
%
We can define groups $\Tor^{A}_{i,j}(\mathbb C,\mathbb C)$, where $i$ is homological grading and $j$ is homogeneous grading.

Consider a space $V^{i,j}=\Tor^A_{i+1,j}(\mathbb C,\mathbb C),\quad i,j\geq 0$.

There is a structure of differential algebra on $T(V)$ with differential of bidegree $(-1,0)$ which is quasiisomorphic to $A$ (see \cite{MSch2} where we heavily used this construction ). According to \cite{Loday} we can compute reduced cyclic homology of $A$ as homology of $\overline{Cyc}(T(V),d)$

Introduce a function $\sum _{i,j\geq 0}(-1)^i dim \Tor^{A}_{i,j}(\mathbb C,\mathbb C)z^j=H(z)$.

Then $A(z)H(z)=1$, hence $H(z)=\frac {1}{A(z)}$( a simple corollary of Bar-duality).

The function $V(-1,z)$ involved in the formula \ref{E:iushbcgh} is equal  
\begin{equation}\label{E:bxvsfg}
V(-1,z)=\sum ^{\infty}_{j=1}\sum ^{\infty}_{i=0}(-1)^{i+1} dim \Tor^{A}_{i,j}(\mathbb C,\mathbb C)z^j=1-H(z)=1-\frac {1}{A(z)}
\end{equation}
After substituting \ref{E:bxvsfg} into \ref{E:iushbcgh} we get our formula.
\end{proof}



Another application is computation of $HC_0(U(TYM))(t)$. This computation was made in \cite{Pol}. It is interesting to compare his and our computations. 

According to \cite{MSch2} the algebra $U(TYM)$ is free and is isomorphic to $T(M)$. The generating function for $M$ was computed in proposition \ref{P:fasdhfjh}. Applying proposition \ref{P:kjcjsh} we get the following result
\begin{proposition}
$\overline{HC}_0(U(TYM))(t)=-\sum_{k>0} ln(1-t^{2k}M(t^{k}))$
where $M(t)$ is given in \ref{P:fasdhfjh}. 
\end{proposition}
\begin{proof}
Is  obvious. The signs in \ref{LLom1} disappear, because $M$ is purely even.
\end{proof}

\end{document}